\documentclass{article}

\usepackage{arxiv}

\usepackage[utf8]{inputenc} 
\usepackage[T1]{fontenc}    
\usepackage{hyperref}       
\usepackage{url}            
\usepackage{booktabs}       
\usepackage{amsfonts}       

\usepackage{microtype}      
\usepackage{lipsum}
\usepackage{graphicx}
\usepackage{longtable}
\usepackage{rotating}
\usepackage[numbers]{natbib}
\usepackage{multirow}
\usepackage[flushleft]{threeparttable}
\usepackage{amssymb}
\usepackage{pdfrender}
\usepackage{tikz,float}
\usepackage{xcolor}
\usetikzlibrary{positioning, shapes.geometric}
\usepackage{enumitem}
\usepackage{array}
\usepackage{lscape}
\usepackage{adjustbox}
\usepackage{tabularx}
\usepackage{svg}
\usepackage{cleveref}
\usepackage{comment}
\usetikzlibrary{backgrounds}

\newcommand{\PreserveBackslash}[1]{\let\temp=\\#1\let\\=\temp}
\newcolumntype{C}[1]{>{\PreserveBackslash\centering}p{#1}}
\newcolumntype{M}[1]{>{\centering\arraybackslash}m{#1}}

\crefformat{section}{\S#2#1#3}
\crefformat{subsection}{\S#2#1#3}
\crefformat{subsubsection}{\S#2#1#3}

\usepackage{tikz}

\usepackage{colortbl}

\graphicspath{ {./images/} }

\newcommand*{\boldcheckmark}{%
	\textpdfrender{
		TextRenderingMode=FillStroke,
		LineWidth=.5pt, 
	}{\checkmark}
}

\newcommand{\pie}[1]{%
	\begin{tikzpicture}[baseline=-2pt]
		\draw (0,0) circle (1ex);\fill[rotate=90] (1ex,0) arc (0:#1:1ex) -- (0,0) -- cycle;
	\end{tikzpicture}%
}

\title{Cybersecurity of Industrial Cyber-Physical Systems: A Review}

\author{
 Hakan Kayan \\
  School of Computer Science and Informatics\\
  Cardiff University, UK \\
  \texttt{kayanh@cardiff.ac.uk} \\
   \And
 Matthew Nunes \\
  School of Computer Science and Informatics\\
  Cardiff University, UK \\
  \texttt{nunesma@cardiff.ac.uk} \\
  \And
  Omer Rana \\
  School of Computer Science and Informatics\\
  Cardiff University, UK \\
  \texttt{ranaof@cardiff.ac.uk} \\
    \And
  Pete Burnap \\
  School of Computer Science and Informatics\\
  Cardiff University, UK \\
  \texttt{burnapp@cardiff.ac.uk} \\
    \And
  Charith Perera \\
  School of Computer Science and Informatics\\
  Cardiff University, UK \\
  \texttt{pererac@cardiff.ac.uk} \\
}
\usepackage{subcaption}

\begin{document}
\maketitle
\begin{abstract}
Industrial cyber-physical systems (ICPSs) manage critical infrastructures by controlling the processes based on the “physics” data gathered by edge sensor networks. Recent innovations in ubiquitous computing and communication technologies have prompted the rapid integration of highly interconnected systems to ICPSs. Hence, the “security by obscurity” principle provided by air-gapping is no longer followed. As the interconnectivity in ICPSs increases, so does the attack surface. Industrial vulnerability assessment reports have shown that a variety of new vulnerabilities have occurred due to this transition while the most common ones are related to weak boundary protection. Although there are existing surveys in this context, very little is mentioned regarding these reports. This paper bridges this gap by defining and reviewing ICPSs from a cybersecurity perspective. In particular, multi-dimensional adaptive attack taxonomy is presented and utilized for evaluating real-life ICPS cyber incidents. We also identify the general shortcomings and highlight the points that cause a gap in existing literature while defining future research directions.

\vspace{0.3cm}


\end{abstract}

\maketitle
\newpage

\section{Introduction}
Industry 4.0 \citenum{lasi_industry_2014} and Industrial Internet \citenum{industrial_internet} have accelerated the integration of industrial cyber-physical systems (ICPSs)  \citenum{rajkumar_cyber-physical_2010} to the various industries ranged from manufacturing \citenum{lee_cyber-physical_2015} to energy management \citenum{parolini_cyber-physical_2010}, water treatment systems \citenum{wang_cyber-physical_2015}, and many more \citenum{yu_smart_2016,chen_intelligent_2017, sampigethaya_aviation_2013}. Critical infrastructures (CIs) \citenum{CIs_public_2019} utilize ICPSs to perform and supervise industrial tasks in harsh industrial environments. Plenty of research \citenum{delsing_migration_2012, wu_cloud_2013, yue_cloud-assisted_2015} has been done on how to facilitate the migration of air-gapped legacy ICS to the modern ones that are even assisted by cloud technologies. These studies are also guided by alliances/institutions such as the Cloud Security Alliance (CSA) \citenum{roza_cloud} to prevent insecure integration. This rapid transformation of ICPSs does bring many cybersecurity challenges. There are three mutually inclusive primary reasons behind these challenges: (i) high connectivity, (ii) increased attack surface, and (iii) heterogeneous infrastructure.

The loss or compromise of CIs that supply fundamental needs may result in catastrophic impact. The survivability of the CIs depends on the security of ICPSs. The defence approaches that focus on CPSs security such as anomaly detection \citenum{schneider_high-performance_2018}, secure routing \citenum{liu_trust_2019}, game theories \citenum{bakker_hypergames_2020, wang_game-theory-based_2016}, and watermarking \citenum{satchidanandan_dynamic_2017}, cannot be directly applied to ICPSs that differ from CPSs in many aspects. The security challenges for ICPSs require unique solutions that consider harsh industrial environments. Even though there is a growing number of publications, the literature that focuses on ICPS security is quite diverse. This independent presentation and evaluation of complementary topics create an unnecessary diversity regarding taxonomies, evaluation metrics, implementation techniques, and test environments. This creates a challenging environment for the new proposals while limiting the discussion of research outputs in a more unified way.

\textbf{Existing Surveys.} The ICPS, ICS, Industrial Internet of Things (IIoT), and Industrial Wireless Sensor Networks (IWSN) are not mutually independent disciplines. Hence, we have also investigated surveys/reviews of them. However, it was challenging to decide research criteria as there is no framework that explicitly describes the relationships of different industrial disciplines studied by the academia. Surveys of ICS challenges from a security point of view based on cybersecurity management are presented in \citenum{cheminod_review_2013, knowles_survey_2015}. Several surveys \citenum{mitchell_survey_2014, humayed_cyber-physical_2017, giraldo_security_2017} classify IDS techniques, review security and privacy issues while systematizing CPSs security and providing a security framework. Surveys of the current challenges for various ICPS architectures are available in \citenum{yu_implementation_2019, leitao_smart_2016, wang_cyber-physical_2008}. The technical review of control engineering tools is presented in \citenum{jbair_industrial_2018} while review of the key enabling technologies and major applications of ICPSs is available in \citenum{lu_cyber_2017}. ICPSs attack detection techniques are surveyed in \citenum{ding_survey_2018, huang_assessing_2018, giraldo_survey_2018,ding_survey_2019, ramotsoela_survey_2018}. Several proposals \citenum{sadeghi_security_2015, xu_survey_2018, liao_industrial_2018, younan_challenges_2020, gidlund_guest_2020, xu_internet_2014} introduce IIoT, study state-of-the-art implementation, and give an outlook on possible solutions while mentioning future research directions. Other surveys \citenum{gungor_industrial_2009, lu_real-time_2016, queiroz_survey_2017} define principles, review technical challenges, and provide structured overview. We classified the previous surveys according to their topics in Table \ref{tab:table1}. We consider all these existing reviews/surveys complementary to our work.

\begin{table*}[ht]
	\centering
	\footnotesize
	\caption{Chronological Summary of Previous Surveys}
	\label{tab:table1}
	\begin{adjustbox}{scale = .8}
		\begin{threeparttable}
			\begin{tabular}{@{}clcccccc@{}}
				\toprule[1pt]
				Year                  & Reference & I & CS & CPS & IoT & WSN & Cybersecurity \\ \midrule[0.5pt]
				$2009$                  & \citet{gungor_industrial_2009}     & \boldcheckmark &    &     &     & \boldcheckmark   &                  \\ \midrule[0.5pt]
				$2013$                  & \citet{cheminod_review_2013}       & \boldcheckmark & \boldcheckmark  &     &     &     & \boldcheckmark   \\ \midrule[0.5pt]
				\multirow{2}{*}{$2014$} & \citet{xu_internet_2014}           & \boldcheckmark &    &     & \boldcheckmark   &     &                  \\ 
				& \citet{mitchell_survey_2014}       &   &    & \boldcheckmark   &     &     & \boldcheckmark                \\ \midrule[0.5pt]
				\multirow{3}{*}{$2015$} & \citet{lu_real-time_2016}          & \boldcheckmark &    &     &     & \boldcheckmark   &                  \\ 
				& \citet{sadeghi_security_2015}      & \boldcheckmark &    &     & \boldcheckmark   &     & \boldcheckmark   \\
				& \citet{knowles_survey_2015}        & \boldcheckmark & \boldcheckmark  &     &     &     & \boldcheckmark   \\ \midrule[0.5pt]
				$2016$                  & \citet{leitao_smart_2016}          & \boldcheckmark &    & \boldcheckmark   &     &     &                  \\ \midrule[0.5pt]
				\multirow{5}{*}{$2017$} & \citet{ding_survey_2018}           & \boldcheckmark &    & \boldcheckmark   &     &     & \boldcheckmark   \\  
				& \citet{lu_cyber_2017}              & \boldcheckmark &    & \boldcheckmark   &     &     &                  \\
				& \citet{queiroz_survey_2017}        & \boldcheckmark &    &     &     & \boldcheckmark   &                  \\
				& \citet{humayed_cyber-physical_2017}&   &    & \boldcheckmark   &     &     & \boldcheckmark                \\
				& \citet{giraldo_security_2017}      &   &    & \boldcheckmark   &     &     & \boldcheckmark                \\ \midrule[0.5pt]
				\multirow{5}{*}{$2018$} & \citet{huang_assessing_2018}       & \boldcheckmark &    & \boldcheckmark   &     &     & \boldcheckmark   \\ 
				& \citet{giraldo_survey_2018}        & \boldcheckmark &    & \boldcheckmark   &     &     & \boldcheckmark   \\
				& \citet{xu_survey_2018}             & \boldcheckmark &    &     & \boldcheckmark   &     & \boldcheckmark   \\
				& \citet{ramotsoela_survey_2018}     & \boldcheckmark &    &     &     & \boldcheckmark   & \boldcheckmark   \\
				& \citet{liao_industrial_2018}       & \boldcheckmark &    &     & \boldcheckmark   &     &                  \\ \midrule[0.5pt]
				\multirow{4}{*}{$2019$} & \citet{ding_survey_2019}           & \boldcheckmark &    & \boldcheckmark   &     &     & \boldcheckmark   \\ 
				& \citet{yu_implementation_2019}     & \boldcheckmark &    & \boldcheckmark   &     &     &                  \\
				& \citet{jbair_industrial_2018}      & \boldcheckmark &    & \boldcheckmark   &     &     &                  \\
				& \citet{younan_challenges_2020}     & \boldcheckmark &    &     & \boldcheckmark   &     & \boldcheckmark   \\ \midrule[0.5pt]
				$2020$                  & \citet{gidlund_guest_2020}         & \boldcheckmark &    &     & \boldcheckmark   &     & \boldcheckmark   \\ \bottomrule[1pt]
			\end{tabular}
			\begin{tablenotes}
				\setlength\labelsep{0pt}
				\footnotesize
				\item $I$: Industrial, $CS$: Control Systems, $CPS$: Cyber-physical Systems, $IoT$: Internet of Things, $WSN$: Wireless Sensor Networks.
			\end{tablenotes}
		\end{threeparttable}
	\end{adjustbox}
\end{table*}

\textbf{Scope of the survey}. The topic of CPS is indeed very popular in academia even though being a fairly new term. This growing popularity is also recognized by the National Institute of Standards and Technology (NIST) that has published a CPS framework \citenum{griffor_framework_2017}. However, although there are similarities, we believe CPS and ICPS security should be treated as a distinct research areas due to attributes (based on the differences between Information Technology (IT) and Operational Technology (OT)) that are unique to industrial environments. Cyber incidents that occur in industrial environments might result in catastrophic failures. Therefore, we have focused on the ICPS security and organized our survey by taking these attributes into consideration.

\textbf{Our contributions.} (i) We briefly define ICPSs, IWSN, IIoT, and ICS by identifying their unique environment characteristics, relationships, and current positions in the literature. We have only investigated and compared the ones that we think as the most similar to ICPSs according to the way that they are studied in the literature. We define the fundamental components of an ICPS and present a modern ICPS architecture (see \cref{section:systemdefinitions}). (ii) We analyze the differences between IT and OT and explain why the security aspect of ICPSs is a unique field that requires particular interest (see \cref{section:ITvsOT}). (iii) We provide a comprehensive review on industrial protocols and infrastructures (see \cref{section:comm.technologies}). (iv) We review the present cyberattack taxonomies from both academic and industrial domains and provide ours which combines their strong aspects (see \cref{section:attacktaxonomy}). Then we present key findings from several industrial reports \citenum{noauthor_2020_2020, noauthor_nccicics-cert_2015, noauthor_ics-cert_2016} (see \cref{section:reports}). Then we analyze the available cyber defense approaches that can be implemented against the top ten most common industrial vulnerabilities (see \cref{section:countermeasures}). (iv) We define ICPS security characteristics (see \cref{section:characteristics}) and (v) review the latest trends on ICPS edge network (see \cref{section:ICPSedge}). (vi) Finally we share the lessons learned (see \cref{section:lessonslearned}) our recommendations (see \cref{section:sec5}) and (see \cref{section:sec6}) conclusions.

\section{INDUSTRIAL SYSTEMS \& INFRASTRUCTURES}
\label{section:systemdefinitions}
\vspace{-0.2cm}
We now briefly define each system, clarify their relationships, introduce ICPS components, and analyze the differences between IT and OT. 
\vspace{-0.1cm}
\subsection{Industrial System Definitions}

\textbf{Industrial Wireless Sensor Networks.} Wireless Sensors Networks (WSN) is made from a group of spatially distributed autonomous/self-processing sensors that simultaneously perform various tasks (e.g., monitoring, detecting, and recording) at a lower cost than wired systems \citenum{pottie_wireless_2000} are deployed in various environments ranging from local (e.g. home, car) to industrial (e.g. military, and health) \citenum{akyildiz_wireless_2002}. In WSN, the current main challenges are related to latency and security \citenum{perrig_security_2004}. These two features are even more significant for IWSN as providing availability is the primary concern for industrial systems deployed to CIs. Thus, various expertise among different disciplines including but not limited to (i) industrial applications, (ii) sensor architectures, (iii) communication/transmission technologies, and (iv) network architectures are desired secure IWSNs \citenum{gungor_industrial_2009}. Current IWSN technologies are supported by standardization organizations (e.g., ISA, IEC, IEEE) due to their adaptability to harsh environments \citenum{lu_real-time_2016}. IWSNs are being integrated into the internet via gateways \citenum{khalil_wireless_2014} and even in the future, they may even have dedicated IPs \citenum{hui_ip_2008}.

\textbf{Industrial Internet of Things.} IIoT refers to a technology emerged from ICS (e.g., Supervisory Control and Data Acquisition (SCADA), Distributed Control System (DCS)) with the integration of interconnected devices, networking architectures (i.e., IWSN), and services through the internet. Conventional manufacturing, automation, computing systems are started to utilize IIoT by implementing cloud infrastructures \citenum{sadeghi_security_2015}. IIoT is composed of two main parts as ICPS: the cyber part (e.g., sensing, networking, computing) and the physical part (e.g., sensors, actuators). Thus, we can identify IIoT as a subsection of ICPS. The defining feature of IIoT is the internet connection provided via network nodes that offer remote management. Besides, the rapid development of cloud technology has made IIoT more attractive as it provides efficient processing and storage of big data \citenum{xu_survey_2018}. The increasing heterogeneity of IIoT has led to the development of WoT \citenum{xu_internet_2014} to solve interoperability problems. We believe in the future, the industrial Web of Things (IWoT) \citenum{jabbar_rest-based_2018} will be a canonical research area as the connectivity grows within the industrial environments.

\textbf{Industrial Control Systems.} ICS is the common term that refers to control systems such as SCADA, DCS, Programmable Logic Controllers (PLC) systems that are located in various industrial sectors. Years ago, ICS was air-gapped while running proprietary protocols therefore being less prone to cyberattacks. Now, ICS adapted to changes that have come with Industry 4.0. Today’s ICS are integrated with IWSN, IIoT, and new generation PLCs that evolves them into advanced ICPSs. SCADA and DCS also are converged together allowing the implementation of hybrid systems that generate new challenges \citenum{kirkpatrick_protecting_2019}. ICS are also targeted by cyberattacks in a daily manner where a successful cyberattack may result in a devastating impact. Therefore, security management \citenum{knowles_survey_2015} is a key point for ICS that is guided by national/regional organizations including the National Institute of Standards and Technology (NIST) \citenum{stouffer_guide}.

\begin{figure}[!h]
	\centering
	\includegraphics[scale = 0.65]{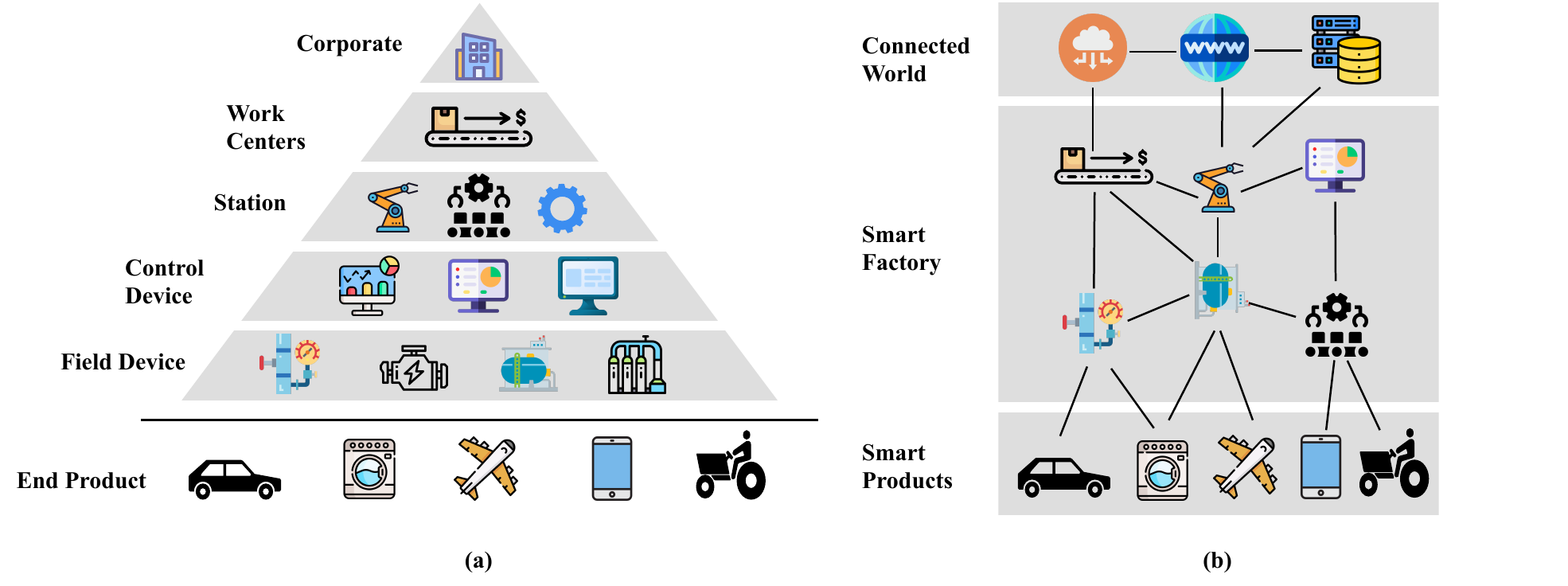}
	\caption{Illustrates the factory hierarchies \citenum{melzer_reference_nodate} of Industry 3.0 (a) and Industry 4.0 (b). The increasing interconnectivity blurred the strict lines between ICPS levels and allowed the development of advanced heterogeneous systems which led to the emergence of the new term “smart factory”.  Increasing overall efficiency is the main motivation behind this transition. However, organizations that tend to rush adopting new technologies without considering how to secure such a transition are prone to cyberattacks.}
	\vspace{-0.3cm}
	\label{fig:fig1}
\end{figure}

\textbf{Industrial Cyber-Physical Systems.} The rapid developments of embedded systems, sensors, and networks resulted in new kinds of mechanisms with multi-tasking capabilities. Thus, the solid line between the physical and cyber environments are started to blur. The term of Cyber-Physical System (CPS) presents today’s advanced computing and networking technologies in a unified way \citenum{lee_cyber_2008}. Industrial Cyber-Physical System (ICPS) refer to CPS that is specifically designed for industrial appliances. ICPSs are deployed to various domains including manufacturing, transportation, healthcare, and energy. The factories belonging to these domains utilize modern ICPSs are called “smart factories”. Figure \ref{fig:fig1} summarizes the difference in hierarchies between old and modern ICPS.

\begin{figure}[!b]
	\centering
	\includegraphics[scale = 0.6]{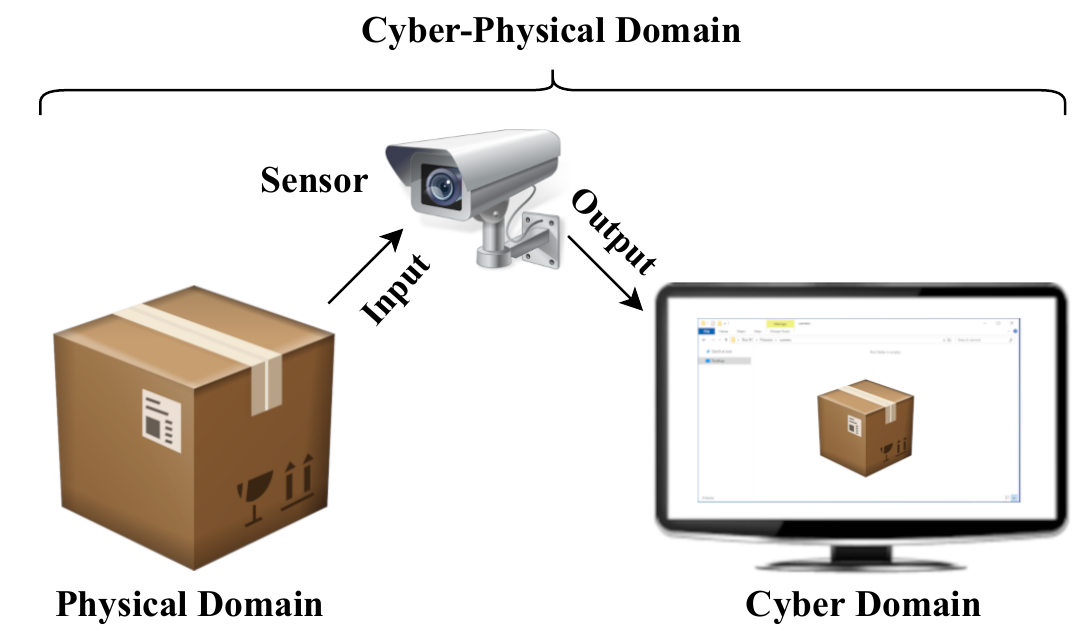}
	\caption{Illustrates the fundamental relationship between cyber and physical domain. To consider a domain as cyber-physical, it should at least include a cyber and physical phenomenon.}
	\label{fig:fig2}
\end{figure}

There are several CPS architectures present in the literature. The most basic one is 3C CPS architecture presented in \citenum{handa_cyber-physical_2015, naoufel_boulila_cyber-physical_2019} where 3C means control, communication, and computation. Other available CPS architectures that classify the CPS domain in a more detailed way are 5C \citenum{lee_cyber-physical_2015, bagheri_cyber-physical_2015}, and 8C \citenum{jiang_improved_2018}. These processes are completed by a variety of interconnected components that are adapting to the latest technologies including real-time multiprocessing. Commercial-of-the-shelf (COTS) products \citenum{noauthor_edge_nodate} that perform the computation on the edge (via edge nodes) to forward already processed data to the cloud to minimize the system load/delay in industrial automation systems are such an example. The main components that are included in most of the ICPS architectures are sensors, actuators, controllers, and Human-Machine Interfaces (HMI). Figure \ref{fig:fig2} introduces the domain relationship of ICPS.

\textbf{Sensors.} Sensors are devices that convert the physical data to cyber data to monitor and forward events to designated components. Their application range varies from smart homes, transportation, manufacturing, medical systems to aviation. They can be grouped either according to their purposes such as temperature, proximity, light, ultrasonic and gas sensors, or their use cases such as industrial, residential, and commercial. As expected, industrial sensors have higher accuracy and durability thus cost higher than their peers while requiring periodic calibration to keep the data integrity. Industrial sensors are mostly utilized for ubiquitous monitoring purposes including human activity \citenum{lin_human_2016, malaise_activity_2018}, gas \citenum{potyrailo_multivariable_2016}, and robotic arm \citenum{kumar_dynamic_2018}.

\textbf{Actuators.} Actuators convert cyber data to a physical phenomenon therefore they are the complementary opposite of sensors. They are usually classified according to their working principles such as hydraulic, mechanical, electric, etc. Recent studies \citenum{lu_real-time_2016, curiac_towards_2016} show that wireless actuators are very promising and even can be utilized for industrial environments where real-time applications take place. However, the new technologies and methods are required to overcome the existing challenges to accept wireless actuators as reliable for performing industrial tasks.

\textbf{Controllers.} The main control unit that gets inputs from sensors and sends outputs to actuators or central units is defined as the controller. The main controller types are Programmable Logic Controller (PLC), Distributed Control Systems (DCS), and Programmable Automation Controller (PAC). Controllers are evolved to the point where they can be utilized interchangeably \citenum{galloway_introduction_2013} with the removal of technological limits that determines their design choices. Additionally, it is even possible to convert a microcontroller such as Raspberry Pi to a PLC via OpenPLC \citenum{alves_openplc_2018} to utilize for a low-cost simulation.

\begin{figure}[!t]
	\centering
	\includegraphics[scale = 0.60]{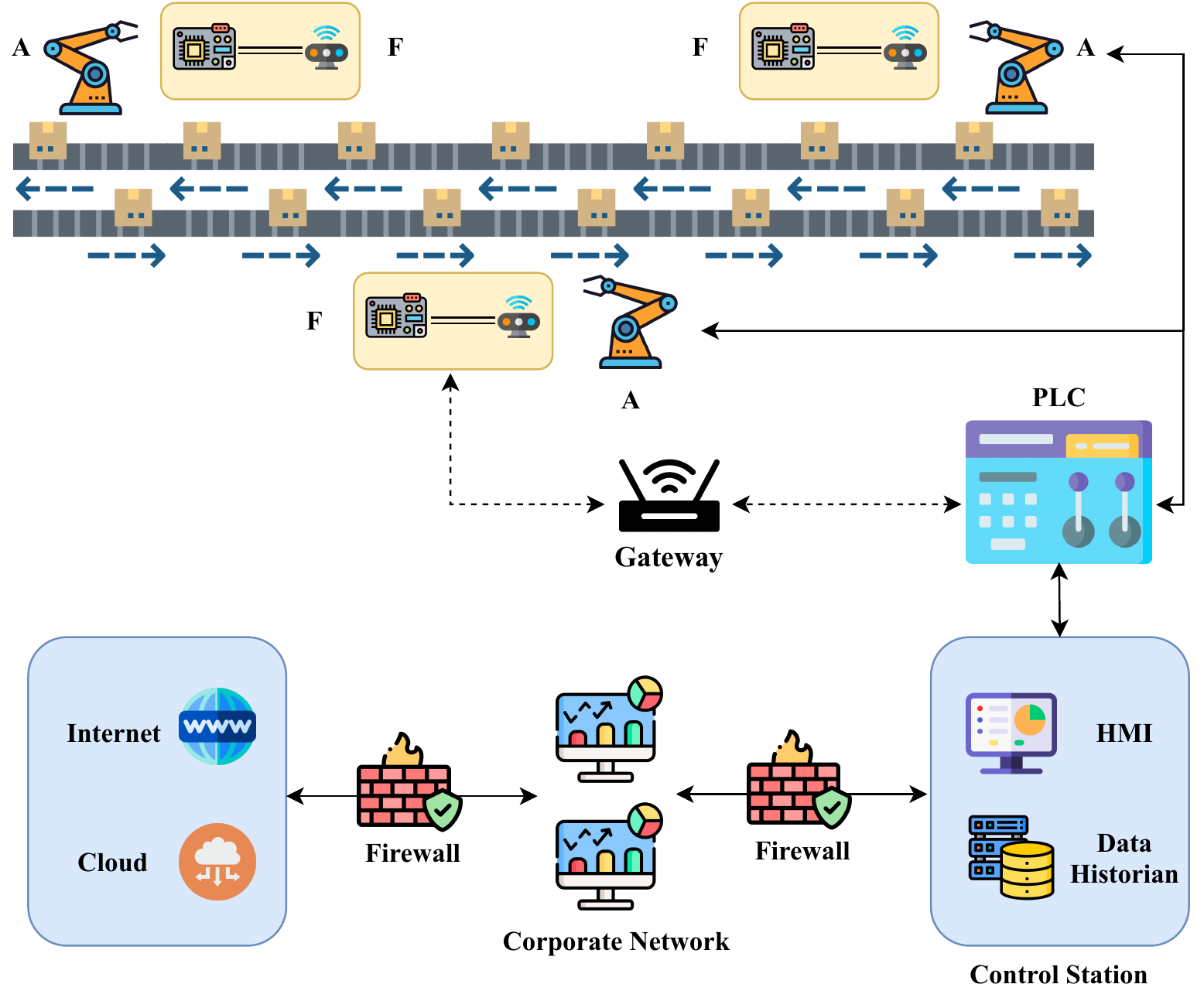}
	\begin{adjustbox}{width=\textwidth,center}
		
		\begin{tabular}{@{}llll@{}}
			$F$: Fog Node & $A$: Actuator & $PLC$: Programmable Logic Controller & $HMI$: Human-machine Interface
		\end{tabular}
	\end{adjustbox}
	\caption{An example of modern manufacturing ICPS architecture. Actuators supervised by fog nodes contain embedded sensors. The data generated by sensors are processed at the edge and sent to PLC over a wireless communication channel. PLC automatically manages actuators based on the upcoming sensor data. Then the data sent to the control station. Here, the data stored in a database are accessible by the corporate network behind a firewall. The same is also applied when connecting outer networks.}
	\vspace{-0.1cm}
	\label{fig:fig3}
\end{figure}

\textbf{HMIs.} Even though fully automated systems are getting popular, ICS that supervise CIs always requires human intervention at some point. HMI is the place where this intervention happens either for monitoring or controlling purposes. HMI technology has already adopted touch screens, and mobile devices. In the future, the cloud-based mobile HMIs \citenum{siddique_controlling_2020} will be more widespread. There are also other ICPS components such as the Remote Terminal Unit (RTU), and data historian that are being integrated into main components to offer high connectivity with simple management. The future ICPSs will be only composed of components with multi-tasking capabilities. The modern manufacturing ICPS architecture contains aforementioned components is presented in Figure \ref{fig:fig3}.

\subsection{The Relationship Between Industrial Technologies}
\label{section:relationships}

We believe the best way to understand the relationship between the aforementioned industrial technologies is to examine the work provided by previous surveys. While \citet{mitchell_survey_2014, leitao_smart_2016, huang_assessing_2018, jbair_industrial_2018} provide a CPS definition, they do not acknowledge its relationship with other technologies. \citet{ding_survey_2018, lu_real-time_2016, yue_cloud-assisted_2015} position ICPS above other industrial terms and claim that IWSNs and industrial wireless sensor-actuator networks (IWSANs) can be referred to as subgroups of ICPSs. \citet{lu_cyber_2017} describes CPS as a technology that integrates the features of IoT and the Web of Things (WoT). \citet{colombo2017} describe ICPS as an industrial discipline that utilizes information communication technologies to integrate digital and physical elements to create intelligent systems. The authors also describe IoT and CPS as complementary disciplines by mentioning that IoT combines ubiquitous systems with CPS technologies. \citet{karnouskos2011} mentions that SCADA relies on CPSs for monitoring purposes, hence defines them as complementary systems. While the most accepted opinion is that the ICPS is the combination of aforementioned disciplines, we could not find any framework that clearly outlines the similarities and differences. We believe using these terms interchangeably results in an independent presentation of complementary works. Thus, there is a need for a framework that positions each term by clarifying their relationships.


\subsection{Information Technology vs Operational Technology}
\label{section:ITvsOT}

We define IT as the technology of processing the information (data) while OT as the technology that focuses on monitoring and controlling physical phenomenon via physical devices and processes. This is why we see IT systems in nearly any kind of environment while we encounter with OT systems in only industrial ones.  OT systems must operate in real-time due to being deployed in such an environment where the availability of the data is the primary concern. The adoption of mobile devices improves the availability of the system by providing ubiquitous access to authorized personnel \citenum{colbert_operational_2016}. Major ICPS suppliers such as Schneider Electric \citenum{papadopoulos_54_2000} and Rockwell Automation \citenum{robert_j_kretschmann_us6167464pdf_2020} already patented such mobile technologies nearly 20 years ago that also confirms how interconnected today’s technologies are.

ICPSs were air-gapped which means they were isolated from any outer networks. Therefore, they were automatically being protected from outsider threats. The only option for an adversary to attack a system was physically inserting data (i.e., via USB sticks) from the inside. Also, the IT and OT domains were kept separated. In today’s ICPS, OT has started to adopt IT-like technologies (e.g., TCP/IP protocol, Windows as an OS) because of its benefits \citenum{adithya_chemudupati_convergence_2012}. However, controlling and monitoring OT devices via IT systems cause new vulnerabilities to appear as the “security through obscurity” approach is no longer applied. The implementation of new security measures requires the harmonization of IT and OT strategies and understanding of basic differences between two technologies.  Table \ref{tab:table2} summarizes the fundamental differences that we consider significant. Relatively new trends such as fog and cloud computing drive further convergence of IT and OT. In addition to these, wireless technology is also now more robust \citenum{noauthor_itot_2018} and deployable for harsh industrial environments. We believe in the future the IT and OT will be integrated, and holistic approaches will form a base for future studies.

\begin{table}[!h]
	\caption{Fundamental Differences Between IT and OT Domains}
	\centering
	\label{tab:table2}
	\footnotesize
	\begin{tabular}{@{}lcc@{}}
		\toprule[1pt]
		& Information Technology (IT)  & Operational Technology (OT)              \\
		\midrule[0.5pt]
		Protocols             & HTTP, TCP/IP, FTP, UDP, SMTP & Modbus, Fieldbus,DNP3, BACnet \\
		Operations            & Stochastic                   & Deterministic                            \\
		Patching(Updating)    & Easy to patch                & Hard to patch                            \\
		Applications          & Time-sharing                 & Real-time                                \\
		Skilled Personnel     & Available                    & Hardly Available                         \\
		Deployment Cost                  & Low                          & High                                     \\
		Security Focus        & Confidentiality              & Availability                             \\
		Authentication Method & Available                    & Barely available                         \\
		Lifecycle             & 3 - 5 Years                  & Over 20 Years                            \\ 
		Communication         & User-centered                 & Machine-centered                        \\
		\bottomrule[1pt] 
	\end{tabular}
\end{table}

\vspace{-0.5cm}
\section{ICPS Communication Technologies \& Protocols}
\label{section:comm.technologies}

We classify the communication protocols according to standard availability, communication type, and network topology. There are two main communication protocol types in terms of standard availability: open, and proprietary protocols. Open protocols may be developed by a single or group of vendors and may require a license fee. They can be utilized with multiple vendors and also supported by a third-party software. On the other hand, proprietary protocols are developed and controlled by a single vendor. They are strictly restricted under legal terms. Legacy industrial systems had proprietary protocols therefore making manufacturer companies dependent on certain vendors. These protocols had been designed to achieve the best efficiency without considering security as a primary concern. The principle of security through obstruction was followed. However, even though there are still such systems, most of the ICPSs are no more air-gapped and even some of them are adopting cloud technologies. This makes open protocols more secure, and popular than the proprietary ones as they are developed by non-profit communities that keep them updated. Also, most companies prefer being more independent \citenum{noauthor_building_2018} when establishing their ICPS as they may integrate their systems with other components when needed.

The features of communication protocols for wireless and wired technologies differ from each other in many aspects. Legacy systems were mostly utilized wired communications technologies, unlike today’s ICPS which have a combination of both. Wired protocols such as Modbus and BACnet also compatible with many wireless technologies including Zigbee \citenum{noauthor_schneider_2015}. Wireless technologies are easy to deploy but may suffer from interference in an environment with high noise like electric distribution facilities. Wired communications are more reliable in terms of speed but harder to install and maintain. The systems that contain both can benefit from the advantages of each technology.

\begin{figure}[!t]
	\centering
	\includegraphics[scale = 0.5]{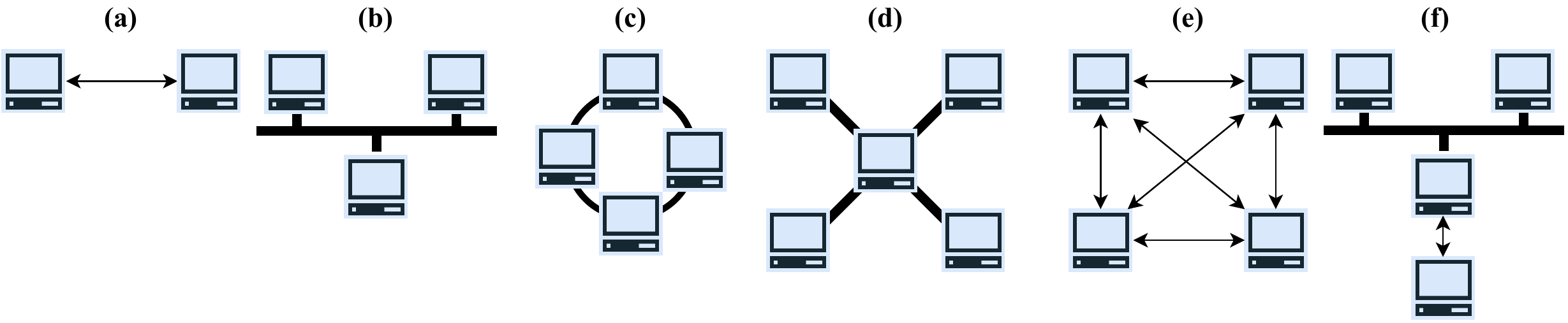}
	\caption{Illustrates industrial communication network topologies in the following order: (a) point-to-point, (b) bus (line), (c) ring, (d) star, (e) mesh, (f) hybrid. Each topology has unique pros and cons. For example, networks with mesh topology are more robust but cost higher, while bus topology costs less but more prone to failures. We can see these kinds of trade-offs for each network topology. Hybrid topologies are preferred in the latest ICPSs due to allowing more interconnectivity.}
	\vspace{-0.6cm}
	\label{fig:fig5}
\end{figure}

Data flow in the communication network may be one-way or bidirectional. Industrial manufacturers/vendors produce a variety of devices that comply with different topologies which determines the arrangement of nodes. These topologies either can be centralized (i.e., star) or decentralized (i.e., mesh). We illustrate the common topologies that are utilized in industrial environments in Figure \ref{fig:fig5}. ICPS can benefit from the high number of nodes in terms of the computational power however it comes with an additional cost. Therefore, minimizing the number of nodes is one of the principles to consider when designing an ICPS. Research that examines this trade-off is presented in \citenum{nedic_network_2018}.

\begin{figure}[!h]
	\centering
	\includegraphics[scale = 0.55]{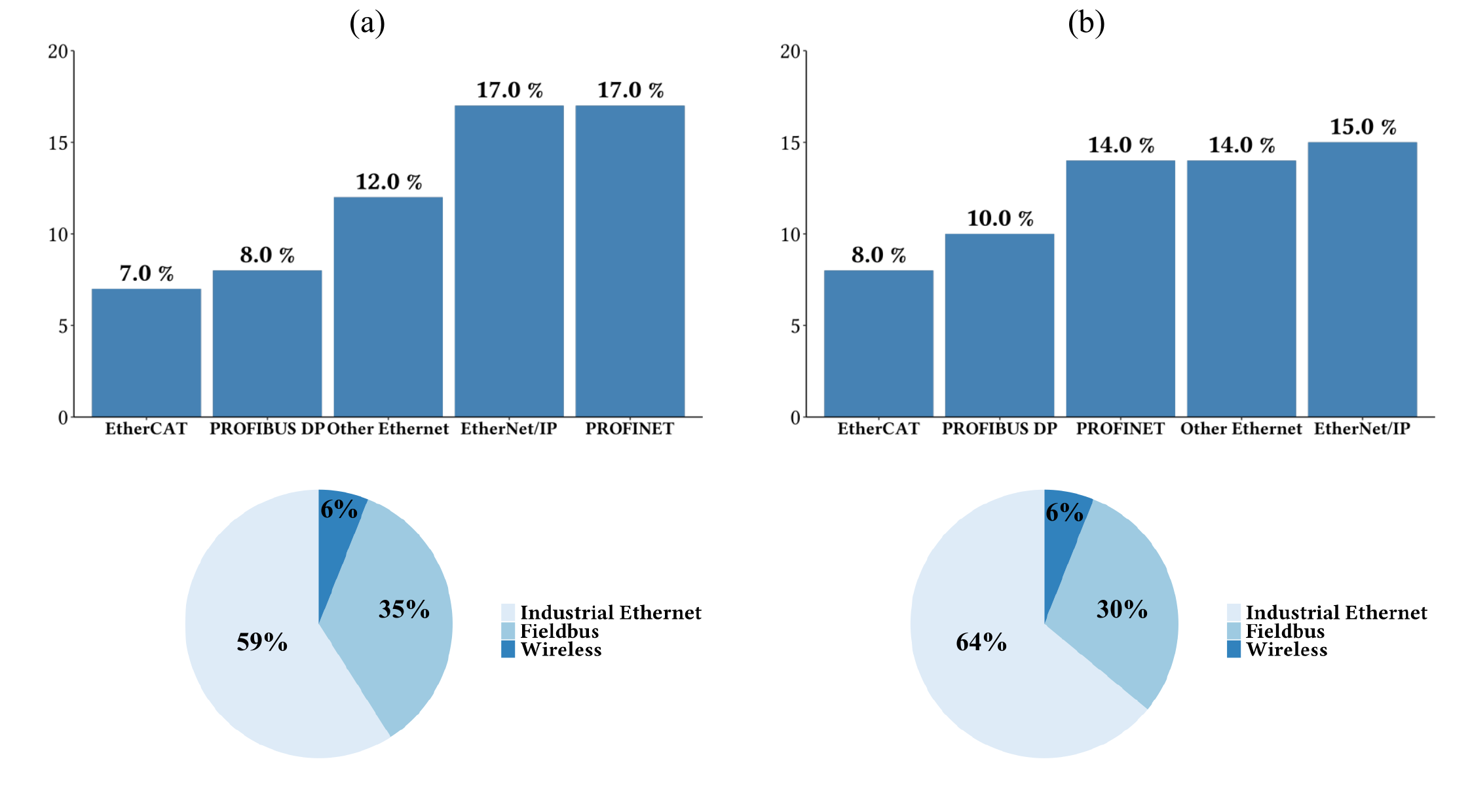}
	\caption{(a) Industrial network shares for top 5 protocols deployed in 2019 \citenum{noauthor_industrial_2019} including how the market shares differ per industrial network technology are shown. EtherNet/IP and Profinet led the industrial network market in 2019. Both of them are industrial ethernet standards while EtherNet/IP is popular in the USA, Profinet is widely accepted in Europe. In terms of overall market shares, Industrial Ethernet has a considerable lead. Increased bandwidth, high data transfer rate, and reliable real-time connection are the main reasons why companies favor industrial ethernet. (b) The same findings for 2020  \citenum{noauthor_industrial_2020} are shown. While the top 5 protocols do not show any significant changes, the Industrial Ethernet deployment rate is 5\% increased while Fieldbus is 5\% decreased which shows there is a transition from Fieldbus to Ethernet.}
	\vspace{-0.3cm}
	\label{fig:fig6_2019}
\end{figure}

We can classify currently available industrial communication technologies according to their working principles: fieldbus \citenum{thomesse_fieldbus_2005}, industrial ethernet \citenum{jasperneite_limits_2007}, and wireless \citenum{li_review_2017}. \textit{Fieldbus}. End-user companies needed a real-time network model where they can connect all their industrial field assets (e.g., sensors, actuators) to increase production efficiency. This led to the development of fieldbus technology. The first ones came as proprietary protocols but later most of the major automation companies established groups/alliances and shared their licenses because the end-users required more heterogeneous production environments. \textit{Industrial Ethernet}. The industrial communication technology started to shift to industrial ethernet from fieldbus because of the promises (i.e., very low latency) that ethernet offered. \textit{Wireless}. The future industrial systems are expected to adopt more wireless technologies because they are easy to deploy and scalable. They also allow a low cost remote management without requiring additional setup. Early wireless systems lacked authentication schemes while having high latency which made them inadequate for industrial systems. Even though these issues are mostly resolved today, sustainability is still a big challenge, as wireless systems operate in resource-constrained environments. Figure \ref{fig:fig6_2019} shows the market shares of deployed industrial network technologies respectively for 2019 and 2020.

\textbf{Fieldbus}. \textit{Modbus}. Open protocol developed and administered by Modbus Organization \citenum{noauthor_modbus_nodate}. It is the most common industrial protocol that has several variants such as Modbus RTU (fieldbus) and Modbus TCP/IP (ethernet). It can be integrated with PLCs ranging from a variety of vendors. \textit{DNP3}. Open protocol owned and developed by the DNP Users Group \citenum{noauthor_dnp3_2020}. IEEE also standardized DNP3 \citenum{noauthor_1815-2012_2020} as it is one of the most common protocols that is utilized in power systems. It does not require additional setup to communicate with non-RTU units which makes it suitable for complex systems. \textit{BACnet}. Open protocol mainly developed for building automation and control networks \citenum{bushby_significant_2002} also utilized in the heating ventilation and air conditioning (HVAC) industry. The protocol is maintained and developed by the American Society of Heating, Refrigerating and Air-Conditioning Engineers (ASHRAE) \citenum{noauthor_bacnet_2020}. Recently, BACnet Secure Connect (BACnet/SC) is proposed by ASHRAE to prevent problems when integrating IT systems into OT infrastructure for cloud-based applications \citenum{david_fisher_bacnet_2019}. \textit{DeviceNet}. Open protocol invented by Rockwell Automation \citenum{noauthor_devicenet_2020}, now developed by ODVA \citenum{noauthor_ODVA_2020}. The reduced number of wires with high flame resistance makes the DeviceNet network suitable for solar applications \citenum{serhane_optimizing_2017}. However, the increased number of nodes in DeviceNet causes an exponential increase in time delay \citenum{deiva_sundari_delay_2017} which may raise an issue for real-time complex networks. 

\textbf{Industrial Ethernet}. \textit{Profinet}. Open Industrial Ethernet Standard developed and administrated by PROFINET International (PI) organization \citenum{noauthor_profinet_2020} that has around 1700 members. It provides a low response time that is ideal for real-time applications. The connection is provided by an ethernet jack that offers high flexibility. Profinet networks can easily interact with the IoT infrastructure \citenum{bellagente_enabling_2016}. \textit{EtherCAT}. Ethernet-based open communication protocol designed for automation to minimize the delay within the industrial network. EtherCAT Technology Group \citenum{noauthor_ethercat_2020} maintains the protocol. EtherCAT has a low implementation cost and also provides high-speed real-time communication \citenum{langlois_ethercat_2018} with low latency \citenum{nguyen_ethercat_2016}. Thus, it is suitable for real-time robotic applications \citenum{sygulla_ethercat-based_2018,delgado_ethercat-based_2016}. \textit{EtherNet/IP}. Open industrial protocol currently maintained and standardized by ODVA \citenum{noauthor_ethernetip_2020}. The EtherNet/IP network can be expanded by ethernet switches that theoretically enables connecting an unlimited number of nodes \citenum{lin_inside_2018}. That also makes EtherNet/IP suitable easy to implement for highly interconnect ICPS. It offers several network structures including ring, star, and linear.

\textbf{Wireless}. \textit{Zigbee}. One of the most popular open wireless communication protocol that is utilized in sectors ranging from industrial automation to smart homes/systems. It is administered and standardized by the Zigbee Alliance \citenum{noauthor_who_2020}. Being a low power solution and allowing remote management makes Zigbee suitable for environmental monitoring \citenum{chehri_zigbee-based_nodate}. \textit{Bluetooth}. Open wireless communication protocol that is designed as Personal Area Network (PAN), managed and standardized by Bluetooth Special Interest Group (SIG) \citenum{noauthor_vision_2020}. Currently, the latest standardized version is Bluetooth 5.0. The rapid development of Bluetooth accelerates its integration into industrial applications. Not supporting mesh topology was a big downside of Bluetooth especially regarding industrial applications and led the proposal of academic solutions as seen in \citenum{seyed_darroudi_bluetooth_2017}. Therefore, Bluetooth SIG added this feature \citenum{darroudi_bluetooth_2019} and standardized it which is considered as a huge step in Bluetooth technology. \textit{LoRaWAN}. Open wireless low-power wide-area network (LPWAN) protocol developed by Semtech maintained and standardized by LoRa Alliance \citenum{noauthor_home_2020}. It is arguably one of the most common LPWAN protocols utilized in many applications ranging from agricultural to home automation. LoRaWAN is suitable for smart home applications but for real-time monitoring, it should only be used where the application does not require a very low response time \citenum{adelantado_understanding_2017}. \textit{WirelessHART}. Open wireless communication protocol based on the IEEE 802.15.4 standard developed by Field COMM Group \citenum{noauthor_product_2016} for industrial control applications. WirelessHART devices have the ability of bidirectional communication that is advantageous for mesh networks. A comprehension of WirelessHART with several wireless standards is presented in \citenum{hassan_application_2017}. \textit{ISA-100.11a}. Open wireless communication protocol based on IEEE 802.15.4 standard developed by ISA \citenum{noauthor_ansiisa-10011a-2011_2020} to provide flexibility for industrial automation systems. ISA-100.11a supports mesh and star topologies. It can also communicate with multiple protocols via gateways thus allowing coexistence of wireless networks. \textit{6LoWPAN}. IPv6 over Wireless Personal Area Networks (6LoWPAN) \citenum{mulligan_6lowpan_2007} is a wireless communication protocol based on IEEE 802.15.4 standard that is developed and standardized by Internet Engineering Task Force (IETF) \citenum{noauthor_ipv6_2019}.  The development of 6LoWPAN was a major stone for IIoT as it has provided an IP-based solution for a range of devices.  However, IP-based solutions rise security concerns due to increased connectivity. We illustrate our security-oriented review of industrial communication protocols in Table \ref{tab:table4}.

\begin{table}[!h]
	\centering
	\caption{Comparison of communication protocols that can be deployed in ICPS. }
	\label{tab:table4}
	\begin{adjustbox}{scale = 0.75, center}
		\begin{threeparttable}
			\begin{tabular}{@{}cccccc@{}}
				\toprule[1pt]
				\textbf{Network   Technology}        & \textbf{Protocol} & \textbf{License} & \textbf{Maintainer}                 & \textbf{Network Topology$\dagger$} & \textbf{Range(meters)$\ddagger$} \\
				\midrule[0.5pt]
				\multirow{4}{*}{Fieldbus}            & Modbus-RTU        & Open$\star$      & Modbus Organization \citenum{noauthor_modbus_nodate}         & Bus/Ring                   & 1500 \\
				& DeviceNet         & Open$\star$      & \citenum{noauthor_ODVA_2020}                                 & Bus/Ring                   & 500  \\
				& BACnet            & Open$\star$      & ASHRAE \citenum{noauthor_bacnet_2020}                        & Bus                        & 1200 \\
				& DNP3              & Open$\star$      & DNP Users Group \citenum{noauthor_dnp3_2020}                 & Bus/Ring/Point-to-Point    & 1200 \\
				\midrule[0.5pt]
				\multirow{4}{*}{Industrial Ethernet} & Modbus-TCP        & Open             & Modbus Organization \citenum{noauthor_modbus_nodate}         & Bus/Star                   & 100  \\
				& Profinet          & Open             & PROFINET International \citenum{noauthor_profinet_2020}      & Bus/Ring/Star              & 100  \\
				& EtherCAT          & Open             & EtherCAT Technology Group \citenum{noauthor_ethercat_2020}   & Bus                        & 100  \\
				& EtherNet/IP       & Open             & ODVA \citenum{noauthor_ODVA_2020}                            & Star/Ring                  & 100  \\
				\midrule[0.5pt]
				\multirow{6}{*}{Wireless}            & ZigBee            & Open             & Zigbee Alliance \citenum{noauthor_who_2020}                  & Mesh                       & 30   \\
				& Bluetooth         & Open             & Bluetooth SIG \citenum{noauthor_vision_2020}                 & Mesh                       & 1k   \\
				& LoRaWAN           & Open             & LoRa Alliance \citenum{noauthor_home_2020}                   & Star                       & 10k  \\
				& WirelessHART      & Open             & FieldCOMM Group \citenum{noauthor_product_2016}              & Mesh                       & 250  \\
				& ISA-100.11a       & Open             & ISA \citenum{noauthor_ansiisa-10011a-2011_2020}              & Star/Mesh                  & 10   \\
				& 6LoWPAN           & Open             & IETF \citenum{noauthor_ipv6_2019}                            & Star/Mesh                  & 100  \\ \bottomrule[1pt]
			\end{tabular}
			\begin{tablenotes}
				\setlength\labelsep{0pt}
				\small
				\item $\star$: The protocol converted to "open" from "proprietary". $\dagger$: Most common network topologies are added, other versions might be available. $\ddagger$: For wired systems "Range" refers to a maximum cable length. Range values are estimated.
			\end{tablenotes}
		\end{threeparttable}
	\end{adjustbox}
\end{table}

\section{ICPS Cybersecurity Analysis}
\label{section:securityanalysis}

ICPS is relatively a new term compared to IWSN, IIoT, ICS, DCS, and SCADA. Even though these systems are complementary to each other, there is an inadequate number of studies that examine the security of them under the ICPS roof. This leads to different study inputs such as taxonomies and evaluation metrics while making it harder to generate a common output for further research. Therefore, we have surveyed the studies that are also related to the these systems and realized there is a need for multi-dimensional adaptive ICPS attack taxonomy. In this section, first we define ICPS attack taxonomy and evaluate real-life ICPS incidents. Then, we summarize key findings from several ICPS vulnerability assessment reports \citenum{noauthor_nccicics-cert_2015, noauthor_ics-cert_2016, noauthor_2020_2020}. Finally, we define ICPS security characteristics and review proposed countermeasures against most common ICPS vulnerabilities.

\subsection{ICPS Attack Taxonomy}

\label{section:attacktaxonomy}

\begin{figure}[!h]
	\centering
	\includegraphics[scale = 0.37]{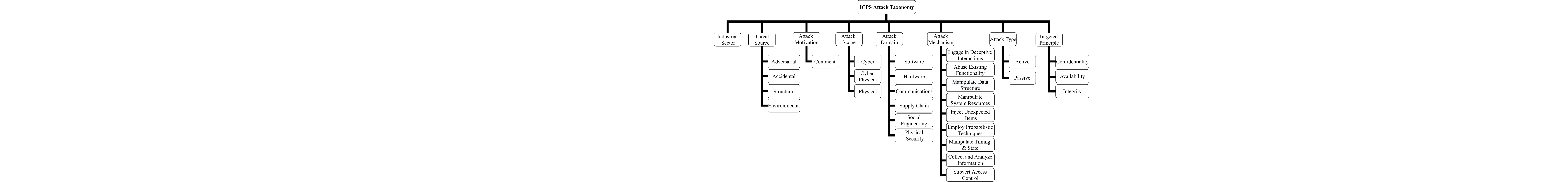}
	\caption{Multi-dimensional adaptive ICPS attack taxonomy. We combine several strong aspects of other taxonomies to build efficient, multi-dimensional, and adaptive ICPS attack taxonomy. "Attack mechanism" and "attack domain" is taken from CAPEC \citenum{noauthor_capec_nodate} while   "industrial sector" is from SIC \citenum{noauthor_nature_nodate}, and "threat source" from NIST \citenum{stouffer_guide}. The only downside of the proposed taxonomy is its dependence on CAPEC. As long as the CAPEC keeps their database up to date, we believe the proposed taxonomy will be sufficient. Even though we claim our taxonomy is suitable and efficient for ICPS attacks in general, we believe application-specific (e.g., manufacturing, transportation) taxonomies have a higher potential to describe industrial attacks more accurately. Only major classes of multi-dimensional branches are shown due to space constraints.}
	\label{fig:fig4}
\end{figure}

According to \citenum{john_d_howard_common_1998}, a well-designed taxonomy should have the following attributes: mutually exclusive, exhaustive, unambiguous, repeatable, accepted, and useful. We believe taxonomies that follow these principles and mention countermeasures and vulnerabilities are more applicable to real-life applications. We have evaluated the available security taxonomies based on their contents and summarized our findings in Table \ref{tab:table3}. We have observed that most of the current taxonomies mainly focus on the IT field and the taxonomies that address OT mostly consider a certain characteristic (e.g., environment, application) which makes them non-usable for different OT systems. We have considered the aforementioned when developing our taxonomy. Also, industrial environments are adapting new ubiquitous technologies hence being more dynamic and heterogeneous. This makes non-adaptive attack taxonomies invalid for future cyber incidents. For this reason, multi-dimensional adaptive taxonomy that is specifically designed for a certain environment is more effective in terms of describing sophisticated cyberattacks. Thus, we have developed such a taxonomy where some of the key features are outsourced to an online attack taxonomy \citenum{noauthor_capec_nodate} that is regularly updated. We now present this taxonomy in Figure \ref{fig:fig4} that contains the following major attributes:

\begin{table}[!b]
	\caption{Classification of the Currently Available Industrial Attack Taxonomies}
	\centering
	\label{tab:table3}
	\footnotesize
	\begin{adjustbox}{scale = .9}
		\begin{threeparttable}
			\begin{tabular}{@{}lccccc@{}}
				\toprule[1pt]
				Reference & IT & OT & Countermeasures & Vulnerabilities & Main Field\\
				\midrule[0.5pt]
				\citet{simmons2014avoidit}               &\pie{360}  &\pie{180}  &\pie{360}  &\pie{180}  &General                \\
				\citet{kim_cyber_2020}                   &\pie{180}  &\pie{360}  &\pie{360}  &\pie{180}  &Nuclear Power Plant    \\
				\citet{loukas_taxonomy_2013}             &\pie{360}  &\pie{180}  &\pie{360}  &\pie{180}  &Emergency Management   \\
				\citet{chapman_taxonomy_nodate}          &\pie{180}  &\pie{180}  &\pie{0}    &\pie{0}    &General                \\
				\citet{hu_taxonomy_2014}                 &\pie{360}  &\pie{180}  &\pie{360}  &\pie{0}    &Smart Grids            \\
				\citet{wu_taxonomy_2017}                 &\pie{180}  &\pie{360}  &\pie{0}    &\pie{180}  &Manufacturing Systems  \\
				\citet{sabillon_cybercriminals_2016}     &\pie{360}  &\pie{180}  &\pie{0}    &\pie{0}    &General                \\
				\citet{brar_cybercrimes_2018}            &\pie{180}  &\pie{180}  &\pie{0}    &\pie{0}    &General                \\
				\citet{applegate_towards_nodate}         &\pie{180}  &\pie{180}  &\pie{360}  &\pie{0}    &General                \\
				\citet{narwal_towards_2019}              &\pie{180}  &\pie{180}  &\pie{0}    &\pie{0}    &General                \\
				\citet{yampolskiy_taxonomy_2013}         &\pie{180}  &\pie{360}  &\pie{0}    &\pie{360}  &CPS                    \\
				\citet{drias_taxonomy_2015}              &\pie{180}  &\pie{360}  &\pie{0}    &\pie{360}  &ICS Protocols          \\
				\citet{berger_attacks_2020}              &\pie{180}  &\pie{360}  &\pie{0}    &\pie{360}  &IIoT                   \\
				\citet{palmer_taxonomy_2009}             &\pie{180}  &\pie{360}  &\pie{0}    &\pie{0}    &DNP3                   \\
				\citet{elhabashy_cyber-physical_2019}    &\pie{180}  &\pie{360}  &\pie{0}    &\pie{0}    &Manufacturing Systems  \\
				\citet{noauthor_capec_nodate}            &\pie{360}  &\pie{360}  &\pie{360}  &\pie{360}  &General        \\        
				\bottomrule[1pt]
			\end{tabular}
			
			\begin{tablenotes}
				\setlength\labelsep{0pt}
				\footnotesize
				\item Legend: \pie{360} : The aspect is explicitly stated and examined, \pie{180} : The aspect is not explicitly stated but partially included by authors, \pie{0} : The aspect is not examined.
				
			\end{tablenotes}
		\end{threeparttable}
	\end{adjustbox}
\end{table}

\begin{enumerate}[label={(\arabic*)}]
	\item \textit{Industrial Sector}: It is significant to define the sector to gain an initial opinion about cyberattacks in general. A food company that operates in two different sectors; food production (manufacturing), and delivery (transportation), may be subject to a cyberattack that targets both. We use the UK’s Standard Industrial Classification (SIC) \citenum{noauthor_nature_nodate} that also complies with the standardization of the European Union \citenum{noauthor_europa_nodate} and the United Nations \citenum{noauthor_unsd_nodate} in our taxonomy.
	\item \textit{Threat Source}: Who or what is behind the incident. It is necessary to identify the threat source to provide further protection for future attacks. In some cases, the threat source might compromise someone else’s cyber source to hide its identity which is a very common case for distributed denial-of-service (DDoS) attacks. In that case, it is significant to distinguish them from victims. We utilize the threat source definition of NIST \citenum{stouffer_guide} in our taxonomy.
	\item \textit{Attack Motivation}: Recognizing the main reason that is directly related to the threat source behind the attack is crucial to determine what you may face in the aftermath of the incident. As the motive can be many things (e.g. financial gain, political opinion), categorizing attack motivation may result in an excessive amount of terms. Therefore, we believe commenting on this issue with a few sentences/words is more practical and informative. 
	\item \textit{Attack Scope}: ICPS is a combination of cyber and physical systems. The attack may target not only the cyber but also the physical domain or both. For example, if the attacker accesses the network and steals data, that attack is cyber only. However, if the attacker gains control of the actuator via unauthorized network access, the attacks become cyber-physical. The only physical attacks are also possible such as physical theft or physically damaging an ICPS equipment. Thus, the scope is divided into three: cyber, physical, and cyber-physical.
	\item \textit{Attack Domain}: Describes the attack pattern. Defining the attacked domain is significant because similar attacks may require similar countermeasures and categorizing them hierarchically supports developing an adequate security plan. We include the multi-dimensional attack domain taxonomy of Common Attack Pattern Enumeration and Classification (CAPEC) \citenum{noauthor_capec_nodate} which is sufficient, detailed, and up to date. There are six main attack domains defined by CAPEC: software, hardware, communications, supply chain, social engineering, and physical security. For example, a common “e-mail injection” attack goes into the following categories in order as follows: software, and parameter injection.
	\item \textit{Attack Mechanism}: It defines the attack technique. Classifying the attack mechanism helps to figure out the vulnerability of the exploited system. One attack may contain several attack mechanisms (techniques) as mostly seen in Advanced Persistent Threat (APT) attacks where the attacker remains undetected for an extended period. We also utilize the multi-dimensional attack mechanism taxonomy of CAPEC \citenum{noauthor_capec_nodate} in this case.
	\item \textit{Attack Type}: The cyberattacks are divided into two categories in general: active, and passive. Imagine an attack scenario where the attacker has gained unauthorized access to the edge data (the data that is driven by sensors) in ICPS. If the attacker just extorts the sensor data without modifying, it is a passive attack. However, if the attacker falsifies the sensor data to further malicious activities the attack turns into an active attack.
	\item \textit{Targeted Principle}: There are three main information security principles: confidentiality, availability, and integrity. Confidentiality refers to the protection of data from unauthorized third parties. Availability refers to the data being accessible by authorized parties whenever needed. Integrity refers to the data being complete and uncorrupted. These three are defined as the CIA triad. One attack may target one or more principles at the same time. For example, while ransomware attacks mostly target availability, malware such as Trojan may target both confidentiality and integrity.
\end{enumerate}


\subsection{Evaluation of Real-life ICPS Incidents Based on ICPS Attack Taxonomy}
\label{section:incidentevaluation}

\begin{figure}[!h]
	\centering
	\begin{adjustbox}{width=\textwidth,center}
		\begin{tikzpicture}[scale = 1.50]

			\foreach \x in {-0.6,1.0,2.6,4.2,5.8,7.4,9,10.6,12.2}{
				
				\filldraw [fill=green!40!blue, draw=none] (\x,0) circle [radius=0.42cm];
				\filldraw [fill=white, draw=none] (\x,0) circle [radius=0.34cm];
			}
			
			\draw (-0.6,0) node [fill=none, anchor = center]{\small 2000};
			\draw (1.0,0) node [fill=none, anchor = center]{\small 2009};        
			\draw (2.6,0) node [fill=none, anchor = center]{\small 2012};
			\draw (4.2,0) node [fill=none, anchor = center]{\small 2013};
			\draw (5.8,0) node [fill=none, anchor = center]{\small 2014};
			\draw (7.4,0) node [fill=none, anchor = center]{\small 2016};
			\draw (9,0) node [fill=none, anchor = center]{\small 2017};
			\draw (10.6,0) node [fill=none, anchor = center]{\small 2018};
			\draw (12.2,0) node [fill=none, anchor = center]{\small 2019};

			\draw (-0.8,-1.0) node [align=left, anchor = center]{\footnotesize 
				\begin{tabular}{lll}
					& Maroochy \citenum{goetz_lessons_2007} \\
					& \\
					&
			\end{tabular}};
			
			\draw (1,0.6) node [align=left]{\footnotesize Stuxnet \citenum{langner_stuxnet_2011}};
			
			\draw (2.6,-1.0) node [align=left, anchor = center]{\footnotesize
				\begin{tabular}{lll}
					&  Aramco \citenum{bronk_cyber_2013} \\
					&  Fukushima \citenum{holt_fukushima_nodate} \\
					&  Niagara \citenum{noauthor_situational_2012}
			\end{tabular}};
			
			\draw (4.2,1.0) node [align=left, anchor = center]
			{\footnotesize
				\begin{tabular}{lll}
					&  Target \citenum{kassner_anatomy_2020} \\
					&  Godzilla \citenum{thornhill_pranksters_2014} \\
					&  Force \citenum{noauthor_ics-cert_2013}
			\end{tabular}};

			\draw (5.7,-1.0) node [align=left]
			{\footnotesize 
				\begin{tabular}{lll}
					& Steel \citenum{lee2014german} \\
					& \\
					&
			\end{tabular}};
			
			\draw (7.2,0.6) node [align=left]
			{\footnotesize 
				\begin{tabular}{lll}
					& Kemuri \citenum{john_leyden_water_2020} \\
					& Ukrainian \citenum{case2016analysis} \\
					&
			\end{tabular}};

			\draw (9.0,-0.61) node [align=left]{\footnotesize TRITON \citenum{aless_triton_nodate}};
			\draw (10.6,0.6) node [align=left]{\footnotesize SCADA \citenum{nj_mahwah_radiflow_2020}};
			
			\draw (12.2,-0.8) node [align=left]
			{\footnotesize 
				\begin{tabular}{ll}
					& Norsk \citenum{leppanen_cyber_2019} \\
					& Riviera \citenum{lindsey_odonnell_post-ransomware_2020}
			\end{tabular}};

			\begin{scope}[on background layer]
				
				\filldraw [fill=green!40!blue, draw=none] (-0.4,-0.06) rectangle (12.61,0.06);
				
			\end{scope}
			
		\end{tikzpicture}
	\end{adjustbox}
	\caption{ The timeline of the significant ICPS attacks.}
	\label{fig:timeline}
\end{figure}
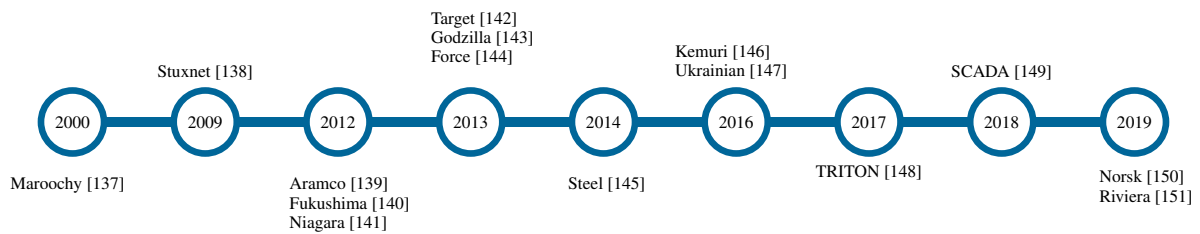

ICPS incidents gain lots of industrial and academic interest as they are discovered. Academia, industry, and even sometimes government entities provide a deep analysis of the incident and publish report/article to inform related communities. Successful attacks with higher impacts subject to more research due to encompassing a variety of aspects including threat actor, attack method, and impact. We have evaluated 15 ICPS incidents (see Table \ref{tab:table6} for evaluation and Figure \ref{fig:timeline} for timeline) based on our multi-dimensional adaptive attack taxonomy.

\textbf{Maroochy Shire Sewage Spill \citenum{goetz_lessons_2007}}. In 2000, a former employee of Maroochy Water Services hacked 142 sewage pumping stations and caused spilling around one million liters of sewage to local water systems. The attack was carried out with just a laptop, compact PC, and radio transmitter. The disgruntled employee accessed the system with stolen system assets and acted as an insider. He actively drove around pumping stations with a car with the hacking tools in it and manipulated the system values. Maroochy Shire Sewage Spill incident is a good example in terms of showing us how using the same credentials that are known by former employees might cause incidents. Industrial organizations must rearrange these credentials to prevent unauthorized access.

\textbf{Stuxnet \citenum{langner_stuxnet_2011}}. In 2009, the nuclear facility of Iran was targeted by the most complex attack known to date. Stuxnet infected more than 100,000 hosts in 25 countries where around 60\% of infected hosts were located in Iran. It specifically targets the vulnerabilities that exist in Microsoft OS and Siemens PLCs while remaining hidden and aiming to generate physical anomalies in CIs. It is claimed \citenum{chen_stuxnet_2010} that the Stuxnet was developed by the US and Israel to sabotage Iran’s uranium enrichment program. However, there was no official confirmation from any sides.

\textbf{Saudi Aramco Attack \citenum{bronk_cyber_2013}}. In 2012, the oil and gas manufacturer Saudi Aramco was targeted by a malware attack later named as Shamoon. Attackers accessed the enterprise network and deployed the malware which deleted data related to production. It is believed that the attackers also had insider help as deploying that kind of malware requires physical access to internal computers. The attack failed to access the industrial network. The hacker group named “The Cutting Sword of Justice” announced that they were behind the attack for political reasons.

\textbf{Fukushima Daiichi Nuclear Disaster \citenum{holt_fukushima_nodate}}. In 2011, the earthquake disrupted Fukushima Daiichi nuclear power station by damaging power systems that cool the reactors which resulted in radioactive contamination and followed by the evacuation of around 100,000 residents. Natural disasters are predictable only to some extent. Usually, the risks that may occur due to these events are ignored when designing ICPS. Fukushima Daiichi nuclear disaster incident led to the re-examination of similar facilities to question how safe they are against natural disasters.

\textbf{Tridium Niagara Framework Attack \citenum{noauthor_situational_2012}}. In 2012, attackers infiltrated the HVAC system of a company located in the USA. The company was using an older version of Tridium Niagara ICS that contains several vulnerabilities that allow backdoor access. These vulnerabilities were already published and analyzed by several cybersecurity organizations. However, the victim company was unaware of the issue. The ICS was connected to the internet with password protection set up. Attackers gained administration privileges without knowing the password by exploiting the already known vulnerability. The attackers did not access any significant document and the motivation behind the attack is unknown.

\textbf{Target Attack \citenum{kassner_anatomy_2020}}. In 2013, the anomalies in the IT system of TARGET were discovered by a third-party forensic team.  The attackers had accessed the personal information of more than 1,000,000 customers. The list of third-party vendors that work with Target was already available in Target’s Supplier Portal. The attackers chose Fazio Mechanical (HVAC manufacturer) and sent a phishing e-mail to one of the employees. They injected malware via a phishing e-mail and stole the login credentials of Fazio Mechanical.  Then, they accessed the enterprise network and stole personal information with the motivation of financial gain. This incident is unique in a way that the attackers accessed the IT network by comprising the OT network first.

\textbf{Godzilla Attack! Turn Back! \citenum{thornhill_pranksters_2014}}. In 2013, the electronic road signs in San Francisco, USA were hacked and instead of essential messages, they were showing “Godzilla Attack!” and “Turn Back!”. The signs were controlled by a third-party company that was still using default credentials to access the network. The incident did not cause any significant problems. The company responsible for the signs claimed that the attack was done by someone who already knew the credentials. It is believed that the motivation behind the attack was just personal entertainment.

\textbf{Brute Force Attacks on Internet-Facing Control Systems \citenum{noauthor_ics-cert_2013}}. In 2013, a gas compressor station owner in the USA alerted ICS-CERT regarding detected increased number of brute force attack attempts on their systems. They were done from 49 different IP addresses. The threat actors and motivation behind the attacks are still unknown.  None of the attacks were successful. This case is a good example to show how early detection of intrusion attempts prevents further damage to industrial systems by increasing the possibility of an effective response.

\textbf{German Steel Mill Cyberattack \citenum{lee2014german}}. In 2014, the blast furnace of a German steel mill was attacked. The attack resulted in massive physical damage since the furnace could not be shut down. Attackers accessed to the enterprise network via phishing e-mail then moved into an industrial network and performed malicious code execution to reprogram several PLCs to compromise the functions of the furnace. They exploited the vulnerabilities resulting from weak boundary protection between established networks due to lack of DMZ. The attack has been defined as an APT and believed to be executed by a group that aims intellectual property theft.

\textbf{Kemuri Water Company Attack \citenum{john_leyden_water_2020}}. In 2016, Verizon published a report of an attack on a water treatment company (Verizon named the company as Kemuri to hide its real identity). The company had stored operational control system credentials on the front-end web server. The adversaries accessed those credentials via SQL injection and phishing. They managed control actuators but failed to cause any harm due to a lack of expertise on SCADA systems.  However, it is believed that the personal information of around 2,500,000 users was stolen.

\textbf{Ukrainian Power Grid Attack \citenum{case2016analysis}}. In 2015, the electricity distribution company Ukrainian Kyivoblenergo was subjected to attack resulted in the power outage that affected 225,000 customers. The attackers accessed the IT network via phishing e-mails, then seized the credentials, and infiltrated the industrial network to execute malware named BlackEnergy 3. They kept attacking the system with 30-minute intervals to prevent mitigation techniques to be deployed.

\textbf{TRITON \citenum{aless_triton_nodate}}. In 2017, the Safety Instrumented System (SIS) of a Middle Eastern oil and gas utility company was shut down due to the successful execution of malware named TRITON. SIS controller that prevents OT assets from malfunctioning made by Schneider Electric SE connected to a Windows PC was deactivated due to TRITON leaving whole utility vulnerable to OT incidents. The attackers first infiltrated the IT network then moved to the OT that shows there was weak boundary protection between these networks.

\textbf{Cryptocurrency Malware Attack on SCADA \citenum{nj_mahwah_radiflow_2020}}. In 2018, a malware was discovered on the OT system of a European water utility company. The malware was designed to mine Monero cryptocurrency by utilizing the HMI and SCADA servers of the victim. It was able to run in stealth mode, but increased CPU and bandwidth usage were detected by the IDS. Updating OT systems requires advanced techniques and thus most of the systems cannot get the latest updates on time. In this case, HMI applications that were not up-to-date were connected to the internet to allow remote management. Attackers exploited these applications to access the system.

\textbf{Norsk Hydro Ransomware Attack \citenum{leppanen_cyber_2019}}. In 2019, aluminum manufacturing company Norsk Hydro suffered from a ransomware attack later named as LockerGoga. The adversaries accessed and encrypted the critical data resulting in the shutdown of the enterprise network and halt of many operations including order processing. The motivation of the attack is estimated as disrupting the production and reputation of the company rather than financial gain as the adversaries chose to execute a previously known ransomware attack after gaining access to the system.

\textbf{Riviera Beach Ransomware Attack \citenum{lindsey_odonnell_post-ransomware_2020}}. In 2019, the water utilities of Riviera Beach (a small city located in Florida, USA) were subjected to a ransomware attack. Attackers sent a phishing e-mail to the police department where the employee opened a malicious link triggered the immediate lockdown of the department computer. The attack spread to all city networks including water utility systems due to being interconnected to the IT network without any bound protection mechanisms. The city council agreed to pay the ransom which was 65 Bitcoins (around \$600,000 due that date) however attackers did not send the decryption key. Then the city council decided to change outdated hardware that was deployed on attacked systems.

Table \ref{tab:table6} demonstrates the evaluation of major ICPS incidents based on the ICPS attack taxonomy that we developed. Our findings have shown that the most targeted industrial sectors \citenum{noauthor_nature_nodate} are manufacturing and electricity, gas, steam and air conditioning supply. Most ICPS attacks are active and organized by nation/state established groups while aiming to disrupt the data integrity. Also, there are no physical-only attacks among major cases.

The attacks against ICPS mostly include more than one stage while the advanced ones may execute the whole CKC. Integrating IT and OT networks without providing robust and secure boundary protection is the most exploited case that has been encountered in ICPS incidents. Attackers mostly target companies that lack security personnel with industrial security expertise via phishing e-mails. Most of the companies that subject to data breach reject publishing a public report on incidents to hide their identities. The information on incidents is mostly available through news agencies or cybersecurity bloggers where they claim getting information via whistle-blowers (e.g., former or current employee with a pseudonym) that makes further examination harder.

\begin{table}[!t]
	\centering
	\caption{Evaluation of Real-Life ICPS Incidents}
	\label{tab:table6}
	\begin{adjustbox}{max width=\textwidth,max height=\textheight}
		\centering
		
		\begin{tabular}{@{}M{1cm}M{3cm}M{1.5cm}M{4.0cm}ccccccc@{}} 
			\toprule[1pt]
			\textbf{Year} & \textbf{Name}& \textbf{Industrial Sector}&\textbf{Threat Source}&\textbf{Attack Motivation}&\textbf{Attack Scope}&\textbf{Attack Domain}& \textbf{Attack Mechanism}&\textbf{Attack Type}&\textbf{Targeted Principle} \\ \midrule[0.5pt]
			2000& Maroochy Shire Sewage Spill& E& Adversarial/Outsider& Revenge& Cyber-Physical& \begin{tabular}[c]{@{}c@{}}Software\\ Communications\end{tabular}& Subvert Access Control& Active & Confidentiality \\ \midrule[0.5pt]
			2009&Stuxnet& C& Adversarial/Nation-State& Sabotage& Cyber-Physical& \begin{tabular}[c]{@{}c@{}}Software\\Hardware\\Communications\end{tabular}& \begin{tabular}[c]{@{}c@{}}Engage in Deceptive Interactions\\Manipulate System Resources\\Inject Unexpected Items\end{tabular} & Active & Integrity\\ \midrule[0.5pt]
			2012& Saudi Aramco Attack& D& Adversarial/Group/Established& Political Reasons & Cyber& \begin{tabular}[c]{@{}c@{}}Software\\Supply Chain\end{tabular}& \begin{tabular}[c]{@{}c@{}}Manipulate Data Structures\\Subvert Access Control\end{tabular}& Active& Integrity\\ \midrule[0.5pt]
			2012& Tridium Niagara Framework Attack& D& Adversarial/Individual& N/A& Cyber& Software& Abuse Existing Functionality& Active& Confidentiality\\ \midrule[0.5pt]
			2012& Fukushima Daiichi Nuclear Disaster& C& Environmental/Natural Disaster & N/A& Physical& N/A& N/A& N/A& N/A\\ \midrule[0.5pt]
			2013& Target Attack& G & Adversarial/Group/Established& Financial Gain& Cyber& \begin{tabular}[c]{@{}c@{}}Software\\Social Engineering\end{tabular}& \begin{tabular}[c]{@{}c@{}}Inject Unexpected Items\\Subvert Access Control\end{tabular}& Active& Confidentiality\\ \midrule[0.5pt]
			2013& Godzilla Attack! Turn Back!& H& Adversarial/Individual& Personal Entertainment& Cyber& Software& Subvert Access Control& Active& Integrity\\ \midrule[0.5pt]
			2013& Brute Force Attacks on Control Systems & D& Adversarial/Outsider& N/A& Cyber& Software& Employ Probabilistic Techniques& Active& Confidentiality\\ \midrule[0.5pt]
			2014& German Steel Mill Cyber Attack& C& Adversarial/Group/Competitor& Theft & Cyber-Physical& \begin{tabular}[c]{@{}c@{}}Social Engineering\\Software\end{tabular}& \begin{tabular}[c]{@{}c@{}}Inject Unexpected Items\\Manipulate System Resources\end{tabular}& Active& Integrity\\ \midrule[0.5pt]
			2016& Kemuri Water Company Attack& E& Adversarial/Nation-State & Sabotage & Cyber-Physical& \begin{tabular}[c]{@{}c@{}}Software\\Social Engineering\end{tabular} & \begin{tabular}[c]{@{}c@{}}Inject Unexpected Items\\Engage in Deceptive Interactions\end{tabular}& Active & Integrity\\ \midrule[0.5pt]
			2016& Ukrainian Power Grid Attack& D& Adversarial/Nation-State & Sabotage & Cyber-Physical& \begin{tabular}[c]{@{}c@{}}Software\\ Hardware\\ Communications\\Supply Chain\end{tabular} & \begin{tabular}[c]{@{}c@{}}Manipulate System Resources\\ Inject Unexpected Items\end{tabular}& Active& Integrity\\ \midrule[0.5pt]
			2017& TRITON& C & Adversarial/Nation-State & Sabotage& Cyber-Physical& \begin{tabular}[c]{@{}c@{}}Software\\Hardware\end{tabular}& \begin{tabular}[c]{@{}c@{}}Inject Unexpected Items\\Manipulate System Resources\end{tabular}& Active& Integrity\\ \midrule[0.5pt]
			2018& Cryptocurrency Malware Attack on SCADA& E & Adversarial/Group/Established & Financial Gain& Cyber& Software& Inject Unexpected Items& Active& Integrity\\ \midrule[0.5pt]
			2019& Norsk Hydro Ransomware Attack& C & Adversarial/Organization& Reputation& Cyber& Software & Inject Unexpected Items & Active& Availability \\ \midrule[0.5pt]
			2019& Riviera Beach Ransomware Attack& E& Adversarial/Group & Financial Gain& Cyber& Software & Inject Unexpected Items & Active& Availability \\ 
			\bottomrule[1pt]
		\end{tabular}
	\end{adjustbox}
\end{table}

\subsection{ICPS Vulnerability Assessment Reports}
\label{section:reports}

ICPSs should be treated as if they will be subject to cyberattack any moment. Security professionals need to discover all vulnerabilities while knowing only one may be enough for an adversary to damage the system. Risk assessment is required to design an efficient security plan. Risk assessment plans vary for each ICPS as they contain different assets that are rapidly evolving due to the integration of new technologies. Such research that considers this change and proposes a risk assessment method for modern smart grids is presented in \citenum{langer_old_2016}.

Industrial Control Systems Cyber Emergency Response Team (ICS-CERT) \citenum{noauthor_industrial_nodate} analyses vulnerabilities of CIs ranged from small and medium-sized businesses (SMBs) to large corporations that are located in the USA. The vulnerabilities are ranked based on The Common Vulnerability Scoring System (CVSS) \citenum{noauthor_common_nodate}. Their industrial vulnerability assessment reports respectively for 2015 \citenum{noauthor_nccicics-cert_2015} and 2016 \citenum{noauthor_ics-cert_2016} outline that the integration of IT to OT causes new vulnerabilities while most of them are related to weak boundary protection. Weak boundaries between OT and IT (enterprise) networks may result in unauthorized access. Establishing an industrial demilitarized zone (IDMZ) \citenum{mazur_defining_2016} is one way to mitigate such a problem. Table \ref{tab:table5} illustrates the key findings from these reports.

\begin{table}[!ht]
	\footnotesize
	\caption{Comparison of ICS-CERT Industrial System Vulnerability Assessment Reports \citenum{noauthor_nccicics-cert_2015, noauthor_ics-cert_2016}}
	\centering
	\label{tab:table5}
	\begin{tabular}{@{}ccccc@{}}
		\toprule[1pt]
		\textbf{Year} & \textbf{Number of Assessments} & \textbf{Found Weaknesses} & \textbf{Weakness Per Asset} & \textbf{Boundary Protection} \\
		\midrule[0.5pt]
		2015          & 112                            & 638                       & 5.7                         & 13.00\%                      \\
		2016          & 130                            & 700                       & 5.4                         & 13.40\%                     \\
		\midrule[0.5pt]
	\end{tabular}
\end{table}

\begin{figure}[!h]
	\centering
	\includegraphics[scale = .65]{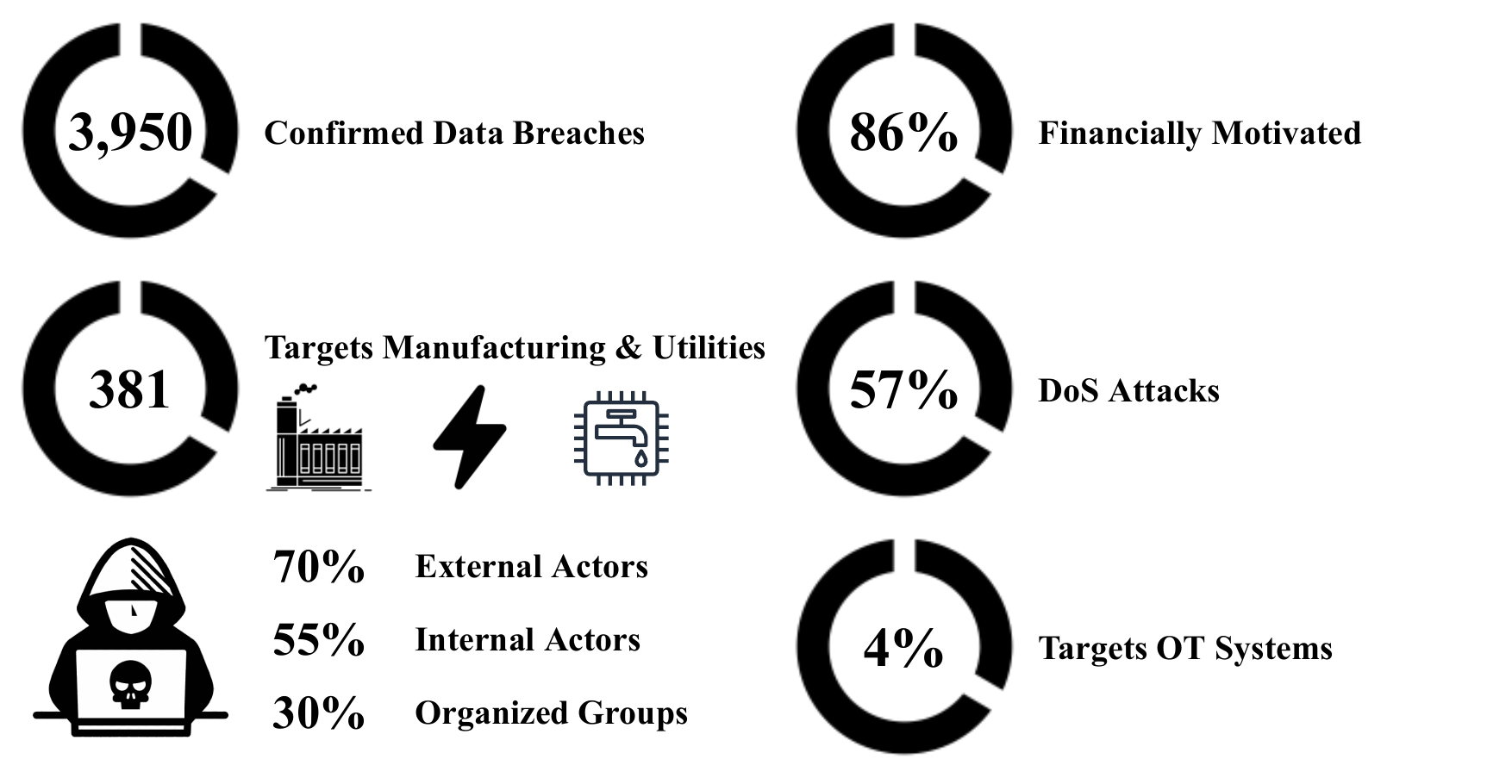}
	\caption{Key findings from Verizon's 2020 data breach report \citenum{noauthor_2020_2020}. 381 data breaches (10\% of total) are against industrial systems where not all of them target OT equipment. Financial gain is the main reason behind 86\% of attacks where most likely they are carried out by organized groups which form 55\% of total threat actors. DoS is the most preferred attack method due to the availability of unsupervised many IoT devices as seen in the Mirai Botnet \citenum{antonakakis_understanding_nodate} incident.}
	\label{fig:verizon}
\end{figure}

Cyber Kill Chain (CKC) \citenum{noauthor_cyber_nodate} is a widely accepted framework created by Lockheed Martin in 2011 that identifies the stages of a successful cyberattack. The CKC developed by SANS \citenum{michael_j_assante_industrial_2015} is more precise and applicable for ICPS in general. Successful delivery of CKC to ICPS results in a data breach that may trigger catastrophic failures. Figure \ref{fig:verizon} illustrates the key findings from Verizon's \citenum{noauthor_2020_2020} data breach report where only 4\% of the total confirmed breaches belonged to the OT systems. However, the results of 4\% may have a bigger impact than the rest (96\% IT-related breaches).

\subsection{Countermeasures Against Most Common ICPS Vulnerabilities}
\label{section:countermeasures}

\begin{figure}[!t]
	\centering
	\includegraphics[scale =.4]{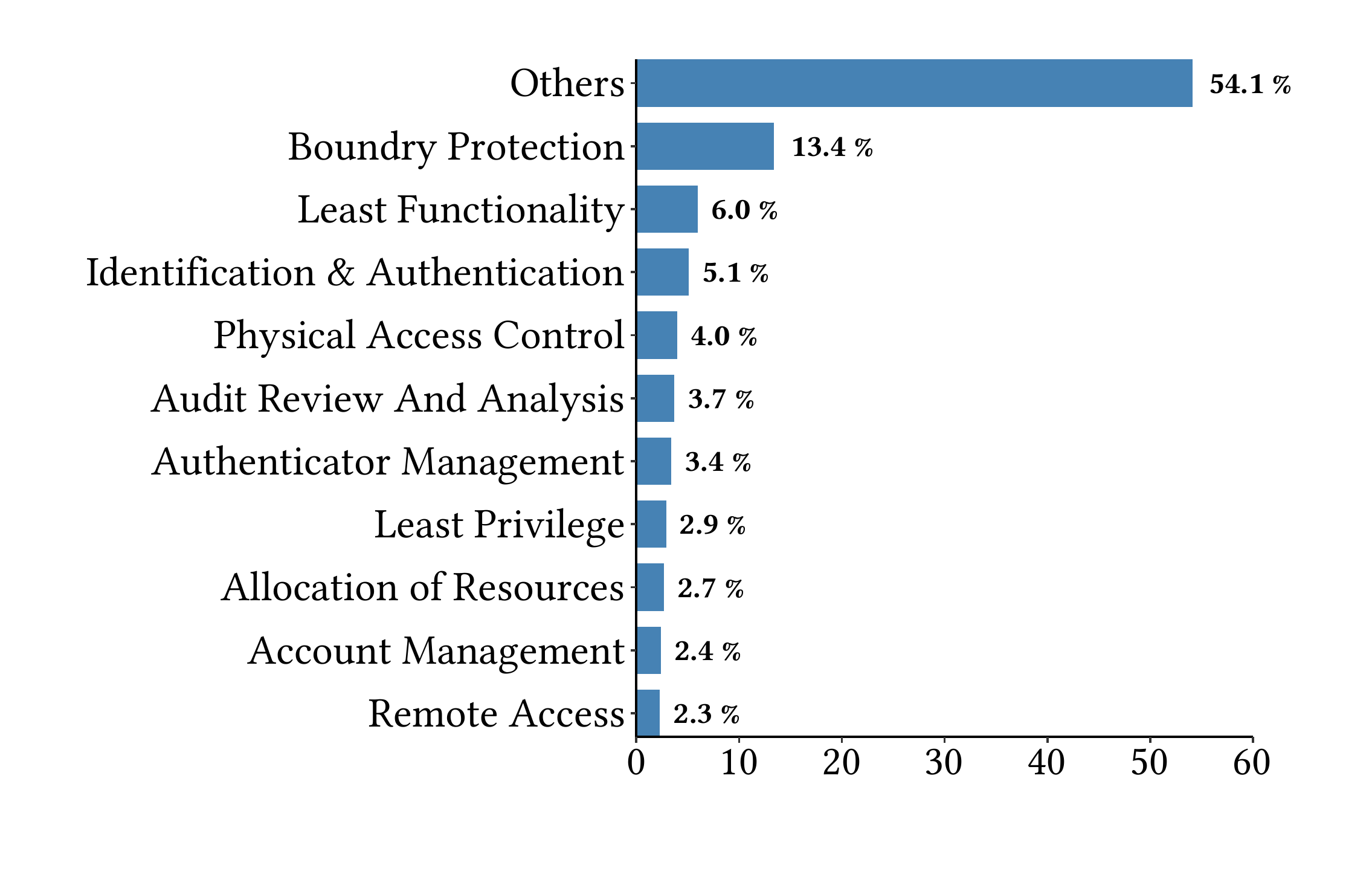}
	\caption{Illustrates the top ten vulnerabilities seen in industrial environments. Vulnerabilities are not isolated from each other. One weakness of ICPS may be a reason for several vulnerabilities. Attacks against CIs target altering the integrity of data first compromise the confidentiality. Therefore, an adversary checks for confidentiality vulnerabilities before engaging in malicious activity. While the current most significant issue is providing boundary protection when integrating new IT technologies to OT, others are mostly present due to a lack of enforcement of well-designed security policies.}
	\label{fig:topten}
\end{figure}

Figure \ref{fig:topten} displays the top ten ICPS vulnerabilities accounting for 45.90\% of the total in 2016 \citenum{noauthor_ics-cert_2016}. While boundary protection is the most common vulnerability, it is followed by weak authentication mechanisms. Integrating IT systems to OT by utilizing the latest available technologies to increase the processing efficiency of factories without properly preparing and complying with security plans, policies, and procedures cause new vulnerabilities that can be exploited by threat actors. The industrial cybersecurity reports \citenum{noauthor_nccicics-cert_2015, noauthor_ics-cert_2016, noauthor_2020_2020} we have examined determine the direction of cybersecurity studies being funded by security companies, research councils, and state establishments. This allows the rapid development of security techniques that provide the deployment of new defence mechanisms. Now, we review academic studies that present such techniques as a solution to the top ten common vulnerabilities (see Figure \ref{fig:topten}) identified by ICS-CERT based on the NIST classification \citenum{stouffer_guide}. We evaluate related studies in the literature based on several characteristics illustrated in Table \ref{tab:tab7} that increase the industrial practicality of these proposals.

\subsubsection{Boundary Protection} 

\label{section:boundary}

The boundary between IT and OT networks in ICPS is fading away due to increased connectivity. Weak boundaries pose a great risk as seen in the past cyber incidents \citenum{lee2014german, john_leyden_water_2020, lindsey_odonnell_post-ransomware_2020}. Establishing a DMZ that contains protection mechanisms including a firewall is the first step to strengthen these boundaries. However, even DMZ itself is prone to attacks. \citenum{korman_analyzing_2017}. \citet{mazur_defining_2016} defines the requirements of a DMZ while discussing if it is needed for mining applications. \citet{jiang_performance_2018} simulate a DMZ using Riverbed Modeler to evaluate performance factors regarding defense-in-depth strategy. \citet{hassan_adaptive_2020} develop IDS based on a semi-supervised deep learning model to provide boundary protection while proposing a framework of attack strategies that target IIoT networks. They evaluate their model on a real-life IIoT network testbed.

\subsubsection{Least Functionality}

The interconnected heterogeneous industrial network contains a variety of system functions (e.g., port, protocol, and services) that keep the main processing units running. Least functionality refers to prohibiting and restricting the usage of these to prevent potential abuse. The industrial companies tend to ignore security due to focusing on reducing cost, hence, resulting in the least functionality vulnerabilities. \citet{cao_security-aware_2020} discuss this issue and propose the random multipath routing model that minimizes the required number of paths to reduce the least functionality vulnerabilities. They utilize that model to provide a security-oriented node deployment framework optimized by a distributed parallel algorithm. Particle swarm optimization (PSO) is another method that addresses optimized node deployment. \citet{ling_quality_2016} propose such an enhanced PSO method that may be utilized for real-world industrial systems to improve efficiency and security aspects.  They apply 17 benchmarking tests to evaluate the proposed model and compare it with previously developed PSO methods while also evaluating their performance on economic load dispatch (ELD) that schedules power generator outputs according to load demands.

\subsubsection{Identification and Authentication}

Applying proper authentication mechanisms in an industrial environment prevents unauthorized access while easing data circulation. Authenticating human-to-machine or machine-to-machine (M2M) communications requires prior identification. Unidentified entities in such environments are prohibited from acting. By faking sensor data, an adversary can damage the ICPS while preventing intrusion detection. Authentication mechanisms are deployed to the edge to prevent such an act. \citet{li_robust_2018} discuss that deploying traditional authentication mechanisms to the resource-constrained environment of sensor nodes poses a challenge and propose a privacy-preserving secure biometrics-based authentication scheme for IIoT. They do the testing via simulation by considering authentication based security properties (e.g., password change, wrong password detection). \citet{esfahani_lightweight_2019} emphasize that and proposes a lightweight authentication scheme for M2M protocols. Their scheme has two steps: each sensor is registered to the system via an authentication server, then mutual authentication is provided. \citet{das_biometrics-based_2018} propose a biometrics-based user authentication scheme for a cloud-based IIoT model that deployed in manufacturing sites while preserving privacy.

\subsubsection{Physical Access Control}

Access to industrial facilities is provided via keys, electronic cards, and mobile technologies. While physical keys are subject to theft, electronic cards and mobile technologies (i.e., smart locks) are prone to forging. \citet{ho_smart_2016} discuss the security of smart locks by modeling with different threat models. They address how current COTS smart locks are vulnerable to state consistency and relay attacks. They propose a touch-based authentication scheme where the communication is provided via bone conduction. \citet{mudholkar_biometrics_2012} briefly introduce biometrics and propose a fingerprint-based authentication mechanism. They claim access to a computer can be provided via fingerprint scanners instead of a password to improve overall security. \citet{mock_real-time_2012} address that the real-time continuous authentication mechanisms provide better security. They develop an iris recognition authentication model based on a commercial eye tracker and claim the current error rate is too high to be used as a standalone mechanism.

\subsubsection{Audit Review and Analysis}

Organizations should adopt Security Information and Event Management (SIEM) system as an operational whole log management mechanism. However, the adversary can organize decoy attacks to occupy SIEM. Thus, prioritizing alerts is an important task. \citet{sancho_new_2020} propose a threat level rating model that evaluates the SIEM output. The authors classify threats and assign four priority levels: critical, important, moderate, and low. First, they perform a data balancing to real-life datasets and then evaluate the proposed model on balanced datasets. They compare the proposed model with commercial software to validate. Due to available limited processing power, an in-depth analysis of ICPS edge data is challenging. \citet{tao_trustdata_2020} discuss these issues and proposes a secure event detection scheme for ICPS while providing a data validation algorithm. They test their system on a real-world dataset and claim that the proposed scheme is adequate to secure data transmission in ICPSs. \citet{huang_next_2014} propose a data historian based on the IBM Informix database for Big Data management.  The authors develop an IIoT data management benchmark (named as IOT-X) by utilizing two relational databases to evaluate their design. They claim the proposed system offers high efficiency compared to traditional historians.

\subsubsection{Authenticator Management}

Three essential principles that must be regulated by password enforcement policies are password change, removal, and encryption. \textit{Password change}. COTS ICPS devices like PLCs come with a default password that is assigned by the vendor. Thus, it poses a great risk and a new password should be set (see \citenum{goetz_lessons_2007} for real-life incident example) before device deployment while being reset at random intervals. \textit{Password removal}. Due to the severity of the operations and availability concerns memorizing a password is not feasible in industrial environments. Therefore, the passwords belong to operational devices are stored in a local/cloud database. When an employee is no longer associated with the organization, the linked password should be removed from the database to prevent possible abuse. \textit{Password encryption}. Strong encryption mechanisms need to be applied to secure passwords starting from the authentication step. If the password is transmitted or kept as plain text, an adversary can access it in case of an intrusion. \citet{sarkar_cyber_2015} address that the password policies for traditional IT networks can be applied to ICPS where the environment contains unique devices such as PLC and RTU. They propose a security-oriented password policy for ICPS with a detailed password creating guideline. \citet{korman_analyzing_2017} evaluate the effectiveness of several ICPS cybersecurity countermeasures including password policy enforcement. Authors analyze the available CPS security assessment tools based on three techniques: network segregation, strengthened access control, and patch management. They claim password policy enforcement complement network segmentation to secure CPSs.

\subsubsection{Least Privilege}

NIST \citenum{stouffer_guide} suggests applying the least privilege principle as a part of the defense-in-depth strategy for ICPS. It provides minimal access to a software/user required for essential tasks when needed, while the task number is being kept at the minimum. This provides easy to analyze systems with higher overall security. Least privilege can be provided via the application of role-based access control (RBAC) that assigns certain access rights to dedicated roles. Such a work is presented in \citenum{kern_using_2018} where authors implement RBAC to the industrial remote maintenance system. They evaluate the proposed system on Raspberry Pi 3 by setting up virtual hosts. They claim such an implementation as a standalone security mechanism can be implemented to remote maintenance systems as a countermeasure against zero-day attacks. Another work \citenum{venugopal_use_2019} evaluates the use of Software-defined networking (SDN) switches to implement the least principle scheme for ICPS. Authors address the cybersecurity requirements on \citenum{stouffer_guide} as a motivation of their work while discussing how to implement NIST suggested mitigation techniques via SDN switches. They test the proposed system via two different SDN switches (real-life testbed) and claim ICPS can benefit in terms of least privilege networking from such an implementation.

\subsubsection{Allocation of Resources}

Due to increasing interconnectivity, providing a secure environment for ICPS is becoming more resource-demanding. \citet{smeraldi_how_2014} discuss the question of how to optimize the cybersecurity budget spending. They apply the knapsack problem and develop an optimization model for two different cases: multiple targets and separate resources, multiple targets and shared resources. They claim such a combinational optimization can be applied for resource allocation. Another model for this issue is presented by \citet{wang_optimal_2017}. The author utilizes mathematical cyber breach probability models to develop a function as a solution to the resource allocation problem. They address that the organizations can benefit if they apply the least privilege principle and access authentication schemes before allocating security resources. \citet{srinidhi_allocation_2015} analyze the resource allocation problem from the manager’s perspective and claim that managers tend to over-invest specific security methods that are effective in the short run due to financial distress that they are faced with while investors prefer more productive ways that are more effective in the long run. They propose a resource allocation decision-support model for managers and investors utilized in case of a breach.

\subsubsection{Account Management}

Additional accounts in ICPS networks are used in exceptional (e.g., error, intrusion) situations. These temporary/emergency accounts may have access to critical assets. Hence, they need to be removed/disabled once their use is completed. Such an action should be regulated by an access control policy. \citet{valenzano} proposes a role-based twofold access control policy model for industrial systems. The author emphasizes the difficulty of validating access policy enforcement. Thus, most solutions in this context either assume that the policy is either enforced or propose an additional software/hardware extensions. The model proposed by the author clearly outlines the usage conditions on edge mechanisms. \citet{ren2019} utilize blockchain technology to implement identity management and access control to edge IIoT mechanisms. The access control policy is defined by the edge network terminal, hence automated. Their evaluation shows that the proposed model can efficiently work in the IIoT edge which is a resource-constrained environment.

\subsubsection{Remote Access}

The integration of cloud technology to OT and rapid advancements in IWSNs have been a huge steppingstone for IIoT and allowed feasible remote monitoring/management within the ICPS. RTUs that gather data from edge sensors are converted to remote substations with the integration of IWSN where the data is accessed and forwarded to a designated point via remote management. However, providing an additional access point increases the attack surface thus new security measures should be set based on security-oriented integration strategy before or when enabling remote access. \citet{alcaraz_security_2013} discuss how to securely integrate IWSN with the internet to provide ubiquitous management for ICPS. Authors address two main challenges: available limited local access options, and the trade-off between real-time performance and security. They analyze the integration strategies and mechanisms from both efficiency and security perspective while addressing internet connection is not required to build remote accessible IWSN. \citet{sadeghi_security_2015} discuss that security has become a hot topic after the integration of IT systems to IIoT that offers remote monitoring and control. They mention future management of IIoT will be challenging due to rapidly increasing heterogeneity that will generate large data. They claim only cloud-based services are capable of processing large data in real-time. However, using cloud services for industrial tasks may raise privacy concerns. \citet{anand_remote_2018} propose a remote water level monitoring design. Authors discuss how current remote access technologies lack modern security mechanisms. They claim existing security measurements for GSM and LTE may be a solution for these issues and can be utilized for remote monitoring while being adapted to IIoT. They also mention that the security of remote access concepts depends on the communication protocol choice.

\begin{table}[!h]
	\caption{The Evaluation of Proposed Countermeasures Against Most Common ICPS Vulnerabilities Based on Dataset Availability, Evaluation Method, Privacy, and Utilization of AI/ML Techniques}
	\label{tab:tab7}
	\begin{adjustbox}{width = \textwidth}
		\begin{threeparttable}
			\begin{tabular}{@{}lcclcccl@{}}
				\toprule[1pt]
				& \multirow{2}{*}{\textbf{ICPS Vulnerability}} & \multicolumn{2}{c}{\textbf{Dataset}} & \multicolumn{2}{c}{\textbf{Evaluated   Method}} & \multirow{2}{*}{\textbf{Privacy}} & \multicolumn{1}{c}{\multirow{2}{*}{\textbf{AI/ML}}} \\ \cmidrule[0.5pt]{3-6}
				&  & \textbf{Pre-obtained} & \multicolumn{1}{c}{\textbf{Generated}} & \textbf{CPS Testbed} & \textbf{Others$\star$} & & \multicolumn{1}{c}{}\\ \midrule[0.5pt]
				\begin{tabular}[c]{@{}l@{}}\citet{mazur_defining_2016}\\ \citet{jiang_performance_2018}\\ \citet{hassan_adaptive_2020}\end{tabular} & Boundary Protection & \boldcheckmark  & & \boldcheckmark  & \boldcheckmark& \multicolumn{1}{l}{}  & \multicolumn{1}{c}{\boldcheckmark} \\ \midrule[0.5pt]
				\begin{tabular}[c]{@{}l@{}}\citet{cao_security-aware_2020}\\ \citet{ling_quality_2016}\end{tabular}     & Least Functionality    & \multicolumn{1}{l}{} && \multicolumn{1}{l}{} & \begin{tabular}[c]{@{}c@{}}\boldcheckmark\\ \boldcheckmark\end{tabular}  & \begin{tabular}[c]{@{}c@{}}\boldcheckmark\\ \boldcheckmark\end{tabular}    &  \\ \midrule[0.5pt]
				\begin{tabular}[c]{@{}l@{}}\citet{li_robust_2018}\\ \citet{esfahani_lightweight_2019}\\ \citet{das_biometrics-based_2018}\end{tabular} & Identification and   Authentication& \multicolumn{1}{l}{} && \multicolumn{1}{l}{}& \begin{tabular}[c]{@{}c@{}}\boldcheckmark\\ \boldcheckmark\\ \boldcheckmark\end{tabular} & \begin{tabular}[c]{@{}c@{}}\boldcheckmark\\ \\ \boldcheckmark\end{tabular} &         \\ \midrule[0.5pt]
				\begin{tabular}[c]{@{}l@{}}\citet{ho_smart_2016}\\ \citet{mudholkar_biometrics_2012}\\ \citet{mock_real-time_2012}\end{tabular} & Physical Access Control & \boldcheckmark && \begin{tabular}[c]{@{}c@{}}\boldcheckmark\\ \\ \boldcheckmark\end{tabular} & & \begin{tabular}[c]{@{}c@{}}\boldcheckmark\\ \boldcheckmark\end{tabular}    & \multicolumn{1}{c}{\boldcheckmark}        \\ \midrule[0.5pt]
				\begin{tabular}[c]{@{}l@{}}\citet{sancho_new_2020}\\ \citet{tao_trustdata_2020}\\ \citet{huang_next_2014}\end{tabular} & Audit Review and Analysis & \begin{tabular}[c]{@{}c@{}}\boldcheckmark\\ \boldcheckmark\\ \boldcheckmark\end{tabular} & \multicolumn{1}{c}{\boldcheckmark}  & \multicolumn{1}{l}{}  & \begin{tabular}[c]{@{}c@{}}\boldcheckmark\\ \boldcheckmark\\ \boldcheckmark\end{tabular} & \boldcheckmark  & \multicolumn{1}{c}{\boldcheckmark} 
				\\ \midrule[0.5pt]
				\begin{tabular}[c]{@{}l@{}}\citet{sarkar_cyber_2015}\\ \citet{korman_analyzing_2017}\end{tabular}     & Authenticator Management    & \multicolumn{1}{l}{} &  & \multicolumn{1}{l}{}& \boldcheckmark& \multicolumn{1}{l}{} &\\ \midrule[0.5pt]
				\begin{tabular}[c]{@{}l@{}}\citet{kern_using_2018}\\ \citet{venugopal_use_2019}\end{tabular}     & Least Privilege & \boldcheckmark &  & \boldcheckmark & \boldcheckmark                             & \multicolumn{1}{l}{} & \\ \midrule[0.5pt]
				\begin{tabular}[c]{@{}l@{}}\citet{smeraldi_how_2014}\\ \citet{wang_optimal_2017}\\ \citet{srinidhi_allocation_2015}\end{tabular} & Allocation of Resources & \multicolumn{1}{l}{} & & \multicolumn{1}{l}{}  & \begin{tabular}[c]{@{}c@{}}\boldcheckmark\\ \boldcheckmark\end{tabular}  & \multicolumn{1}{l}{} &\\ \midrule[0.5pt]
				\begin{tabular}[c]{@{}l@{}}\citet{valenzano}\\ \citet{ren2019}\\ \end{tabular} & Account Management & \multicolumn{1}{l}{} && \multicolumn{1}{l}{}  & \begin{tabular}[c]{@{}c@{}}\boldcheckmark\\ \boldcheckmark\end{tabular} & \begin{tabular}[c]{@{}c@{}} \\ \boldcheckmark\end{tabular} & \\ \midrule[0.5pt]
				\begin{tabular}[c]{@{}l@{}}\citet{alcaraz_security_2013}\\ \citet{sadeghi_security_2015}\\ \citet{anand_remote_2018}\end{tabular} & Remote Access                                & \multicolumn{1}{l}{} & & \boldcheckmark & \multicolumn{1}{l}{} & \begin{tabular}[c]{@{}c@{}}\boldcheckmark\\ \boldcheckmark\end{tabular} & \\ \bottomrule[1pt]                                                    
			\end{tabular}
			
			\begin{tablenotes}
				\setlength\labelsep{0pt}
				\item  $\star$: Others include evaluation via benchmarking tests, simulations, or proof of concept.
			\end{tablenotes}
		\end{threeparttable}
	\end{adjustbox}
\end{table}

\subsection{ICPS Cybersecurity Characteristics}
\label{section:characteristics}

The most sophisticated cyber attacks (e.g., Stuxnet, TRITON) in history targets CIs that are managed by ICPS. Therefore, the concept of defense-in-depth must be applied to all assets contained in CIs.  Defense-in-depth can be provided via establishing several defense layers where each layer serves for a certain purpose while the main objective is to provide a secure environment. These layers may differ for technical assets (e.g., hardware, software) while showing similarities in terms of personnel and procedures. Secure industrial environment that is designed based on defense-in-depth approach should contain the following characteristics that complement each other: 

\textit{Robustness}. All systems are prone to fail. Robustness determines how much a system can endure before failing. This is a significant security feature for each asset in industrial environments due to the cascading effect observed in highly interconnected ICPS. The robustness of a system should be tested whenever a change is made to the system. Even though there are not any known changes, periodically testing is required as some of the components may degrade over time. Fuzz testing \citenum{voyiatzis_modbustcp_2015} such as Netflix’s Simian Army approach \citenum{tseitlin_antifragile_nodate} can be implemented to evaluate ICPS robustness.

\textit{Resilience}. We can shut down IT systems whenever an anomaly is detected. However, this is not valid for OT systems as they are supervising CIs, they need to be kept operating even when there is an intrusion. Resilience determines how long does it take for the system to fully recover after an anomaly. SDN \citenum{babiceanu_cyber_2019, al-rubaye_industrial_2019} is one of the techniques that may be utilized to develop models that contain routing algorithms to increase resiliency in ICPSs.

\textit{Redundancy}. If we can answer the question of what happens when one sensor fails during the manufacturing process with the back-up sensor activates and keeps reporting, that means we designed a redundant system. The edge monitoring mechanism of ICPS consists of a sensor that supervises production line assets such as robotic arm, conveyor belt, gas tank, and oven. The data taken from these are either sent to the command and control (C\&C) center to be checked by the control engineer or handled by autonomous control units. The final phase of the attack against ICPS includes faking sensor values to delay the activation of SIS. This can be prevented via deploying additional layers of sensors under the context of increasing redundancy. Therefore, even though it is not directly mentioned when designing optimized systems, redundancy is one of the main factors that is considered. Such an example is presented in \citenum{karabulut_optimal_2017}.


\subsection{Securing ICPS Edge Network}
\label{section:ICPSedge}

The real-life ICPS incident evaluation (see \cref{section:incidentevaluation}) has shown that most adversaries target edge network/mechanisms by either exploiting weak boundary protection mechanisms (i.e., accessing OT assets from IT domain) (see \cref{section:boundary}) or infiltrating other ICPS elements (e.g., HMIs, PLCs) as the main motivation behind the attacks to maximize given damage by disrupting actuator behaviours \citenum{goetz_lessons_2007, langner_stuxnet_2011, lee2014german, aless_triton_nodate}. Therefore, their first step after breaching the system is to fake sensor readings to bypass deployed anomaly detection mechanisms. Thus, even though the vulnerability assessment reports \citenum{noauthor_nccicics-cert_2015, noauthor_ics-cert_2016, noauthor_2020_2020} show that the most common ICPS vulnerabilities based on weak boundary protection mechanisms, as the final aim of the adversary to disrupt ICPS edge network, efficient security mechanisms that feature key cybersecurity characteristics (see \cref{section:characteristics} and detect the anomalies in physical behaviours (e.g., change in the temperature, pressure, fan behaviour) should be deployed to the edge. The evaluation based on real-life testbeds and datasets generate the most realistic results for such research. Now we briefly summarize the latest research related to ICPS edge security based on the findings of previous surveys/works (see Table \ref{tab:table1}).

\textit{Edge Anomaly Detection}. \citet{giraldo_survey_2018} survey the physics-based attack detection techniques in cyber-physical systems and propose a taxonomy to evaluate related research. Their key findings include: (i) the vast majority of papers do not share common evaluation metrics and do not simultaneously utilize simulation, testbed, and real-world data, (ii) the cases when adversaries in control are ignored. The authors emphasize that the anomaly detection monitor should be deployed to the edge and not just to the central network while proposing new evaluation metrics that can be applied to a variety of anomaly detection algorithms. \citet{ramotsoela_survey_2018} survey the anomaly detection methods utilized for IWSNs. The authors mention that the trade-off between detection accuracy and power consumption is one of the main issues to be considered while the other one is the lack of training data. They emphasize the high cost and inability to evaluate complex ML algorithms are the main drawbacks of IWSN testbeds. Besides, many papers utilize simulation programs, hence, we can conclude that the access to these testbeds is also questionable. \citet{shah2018anomaly} apply anomaly detection by utilizing several machine learning techniques on data gathered from real-life industrial machines. The authors conclude that while in some use cases the anomalies can be detected via statistical analysis, others require machine learning techniques. They mention the data deviation due to external reasons (e.g., at the start, malfunctioning, being idle, degradation) occurs more than expected so should be considered when training the ML model.

\textit{ICPS Testbeds}. \citet{mclaughlin2016cybersecurity} summarize the required features of an efficient ICPS testbed and emphasize that the hardware is a must for an ICPS testbed, hence, hardware-in-the-loop (HIL) testbeds are better at simulating real-world cases. The authors also mention that the HIL testbeds are becoming standard for vulnerability assessment thanks to their increasing numbers and allowing the testing of cyber-physical components. \citet{yamin2020cyber}
propose an extensive survey regarding security testbeds. The authors confirm that the interest in security testbeds (emulation, simulation, hybrid, or real) is increasing. However, the efficiency of these testbeds is questionable due to the lack of quantitative and qualitative analysis. \citet{holm2015survey} review 30 ICS testbeds based on the evaluation metrics defined by \citet{siaterlis2012use} which are \textit{fidelity, repeatability, measurement accuracy}, and \textit{safe execution of tests}. The authors describe fidelity is the most key characteristic as it defines the accuracy of the testbed. However, only 4 of the works discuss the fidelity of the testbeds based on the standards \citenum{stouffer_guide}. Also, the objectives of the testbeds are defined in detail. Hence, the authors emphasize the need for a comprehensive evaluation framework to be utilized to compare the available ICS testbeds.

\textit{ICPS Datasets}. \citet{mitchell_survey_2014} survey 28 papers that propose IDS for CPS based on the detection technique and audit material. 24 out of 28 papers utilize datasets while 22 of them do not share them. 6 papers use simulated datasets rather than operational ones. The authors also define physical process monitoring is one of the key aspects of intrusion detection. \citet{khraisat2019survey} review the IDS datasets. The authors mention that the utilization of older datasets accepted as benchmarks results in inaccurate claims due to their lack of current sophisticated malware. Thus, there is a need for an up-to-date publicly available dataset. \citet{ahmed2016survey} also propose such a survey. The authors claim that the real reason behind the lack of publicly available datasets is privacy-related issues. They also find the usage of older datasets problematic. They emphasize the importance of dual optimization to achieve simpler datasets. \citet{zolanvari2018effect} study the place of ML techniques in IIoT. The authors emphasize that the anomalies correspond to around 1\% of the total data in real-life cases, hence causing the generation of imbalanced datasets. However, training ML models via imbalanced datasets cause generation of inaccurate security mechanisms and the techniques (e.g., oversampling, undersampling) used to overcome this issue have their drawbacks.

The utilization of ML techniques to detect anomalies in industrial systems is favored by academia. However, there are still many challenges to be addressed including operating in resource-constrained environments, and dealing with anomalies resulted from non-adversary events. The increase in the number of testbeds (e.g, physical, simulation, HIL, emulation and virtualisation) is another positive development, however the access to cyber-physical testbeds that provide the most realistic results is questionable as more research is done via simulation only testbeds. The older datasets that are accepted as benchmarks are still widely used even though they do not present the current cybersecurity environment. Privacy is the main concern behind the lack of up-to-date public datasets. Besides, evaluating works that do not release the utilized dataset is more challenging. 

\section{LESSONS LEARNED}
\label{section:lessonslearned}

Dissimilar to IT security, industrial security is still in early development and gradually adapting to scientific research. Integration of IT systems and the rapid development of technologies based on ubiquitous computing (e.g., IIoT, IWSN) that increases the overall efficiency, heterogeneity, and interconnectivity of industrial systems makes the currently available security measures inadequate. We make the following observations based on our review: (i) the relationship between emerging new industrial technologies requires clarification, (ii) industrial cybersecurity policy based solutions lack common evaluation framework, (iii) non-adaptive cybersecurity solutions lose validity over time, (iv) security policies and redundant solutions are overlooked, (v) inefficient realistic testbed and up-to-date dataset utilization. We now examine these in detail.

\textit{Confusion over lack of classification}. New terms emerge from new technologies. To prevent confusion and overlapping, we need to clarify the relationship of such terms (e.g., ICS, IIoT, IWSN, WSAN, SCADA, DCS, and IWoT) with the other complementary industrial disciplines. We only have defined the ICS, IIoT, and IWSN other than ICPS because we find relevant studies in the literature. However, we have not provided a clear distinction as it requires further study. The classification framework that explicitly states the relationship of terms that are used to define systems located in industrial environments will prevent the diversion of complementary future industrial research. 

\textit{Industrial cybersecurity policy-based solutions lack common evaluation ground and framework}. While the top industrial cybersecurity vulnerabilities occur due to weak boundary protection, they are followed by security policy-based weaknesses (e.g., least functionality, identification and authentication, physical access control, authenticator management, least privilege) that can be evaluated under the access control policies. Implementing such a policy on a real ICPS environment while simulating attacks and observing for a certain period may be the most realistic way to evaluate, however, disrupting an ICPS is not acceptable due to supervision of critical tasks. Several proposed solutions apply hardware extensions to enforce these policy-based models but the efficiency of these methods is questionable due to the lack of an evaluation framework.

\textit{Non-adaptive cybersecurity solutions/taxonomies lose validity}. Legacy air-gapped ICPS had a static structure that was built from components expected to work at least 15 years without any significant changes. However, current ICPS are dynamic due to constantly adopting new components/technologies. Besides, suitable components (e.g., PLC, HMI) are updated when a vulnerability is discovered to prevent further abuse. Our real-life incident evaluation has shown that outdated security mechanisms are the main reason for the data breach. However, redeploying is not feasible as they are obliged to non-stop monitor critical tasks. Thus, we need flexible, adaptive solutions that can continue to operate with minimum human intervention. Such solutions can only be produced via utilizing testbed, dataset, and machine learning algorithms.

\textit{Inefficient realistic testbed and up-to-date dataset utilization.} Even though there are many realistic testbeds available, the evaluation that shows how close they mimic their real-life counterparts is hard to find. Besides, the vast majority of research that focus on ICPS, and CPS anomaly detection prefer simulation only testbeds and the evaluation phase mostly contain only one type of testbed rather than combining several ones. Hence, we can question the accessibility of these realistic testbeds. The issue regarding datasets is the lack of publicly available ones that include network traffic containing latest malware. Also, many research do not release the dataset they utilize, hence, making it hard to compare with similar works.  

\textit{A lack of security policy studies and redundant solutions}. Humans are the weakest link in the information security chain. Enforcing security policies is the most feasible solution to prevent human-centric errors/misuses. However, our study has shown while academia heavily favors developing intrusion detection systems (post-attack), it lacks in terms of security policy based studies (pre-attack). On the other hand, redundant systems are also overlooked. We have realized while most papers focus on resilience and robustness while slightly mentioning redundancy. 


\section{Research Challenges and Directions}
\label{section:sec5}

\begin{figure}[!t]
	\centering
	\includegraphics[scale = .55]{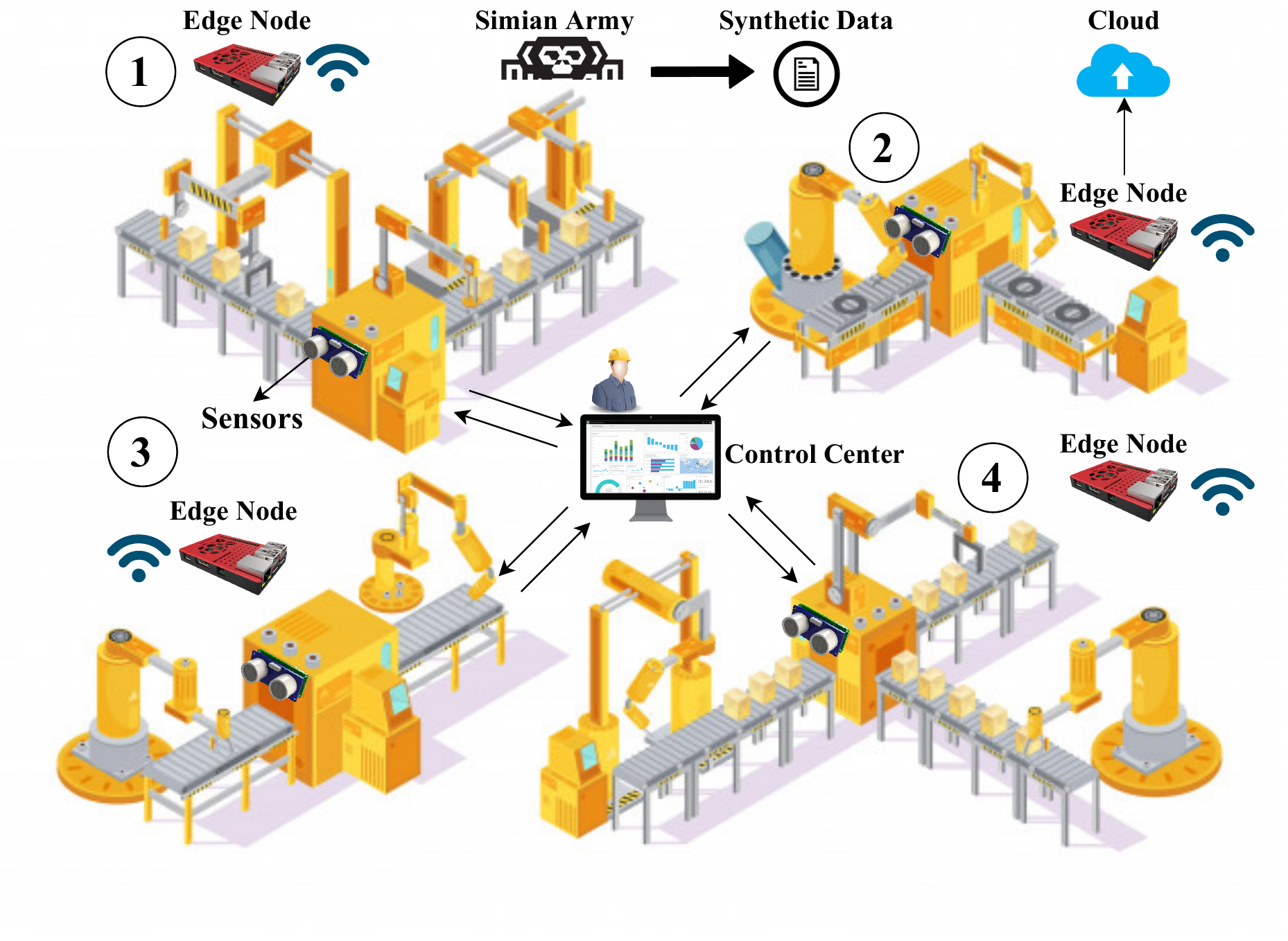}
	\caption{Example of an ideal ICPS evaluation environment. Four testbeds with different contexts are present. Each testbed is supervised by context-aware sensors connected to the main network and edge nodes deployed on an isolated network connected to the cloud. The control center monitors and manages each testbed. Simian Army \citenum{tseitlin_antifragile_nodate} approach is applied to conduct attacks so we can generate a robust synthetic dataset that contains mixed network traffic. The adaptability is present thanks to context-aware sensors/edge nodes while edge notes also provide redundancy.}
	\label{fig:figRC}
\end{figure}

In this section, we identify the research challenges derived from the evaluation of this survey. Our findings show that there are challenges in the field of ICPS cybersecurity that are based on the lack of adequate evaluation/test environments that utilize up-to-date datasets, variety of testbeds while adapting unified evaluation methods. Thus, novel techniques should be employed to provide adequate solutions for these unique challenges where the "uniqueness" comes from being in an "industrial" environment adopting recent ubiquitous computing and communication technologies. We illustrate an ideal ICPS evaluation environment as a solution to these challenges that are based on the future directions derived from our survey in Figure \ref{fig:figRC}.

\subsection{Adaptability and Context Awareness}

Cybersecurity is now one of the most dynamic disciplines due to the high interest of large entities (i.e., states/nations) in cybercrime. The vulnerability assessments reports we analyzed have shown that ICPS cyber incidents (as similar in \citenum{nj_mahwah_radiflow_2020}) occur due to the utilization of out of date security mechanisms. These non-adaptive cybersecurity mechanisms are prone to fail before attacks utilizing the newest methods (e.g., zero-day attacks, APTs). ICPSs that consist of continuous processes require real-time supervision (by sensors) without human intervention. Hence, the deployed cybersecurity mechanisms should provide adaptive, autonomous, and non-stop protection. These challenges are also valid for cyberattack taxonomies. Hence, we have proposed an adaptive ICPS attack taxonomy. However, the validity of our taxonomy depends on CAPEC \citenum{noauthor_capec_nodate} and it may not reflect all characteristics of specific industrial applications.

Context-awareness \citenum{perera_context_2014} is a promising feature that we expect to see in further adaptive ICPS edge security related studies. The AI/ML algorithms are utilized to develop ML models that consist several steps including data gathering, parsing, and training. Current ML models/workflows are automated via machine learning pipelines based on open-source frameworks or cloud ML services to simulate interconnected environment and achieve realistic results. Due to carrying out these processes in an accessible cyber environment, these models are prone to adversarial machine learning techniques \citenum{qian2020orchestrating} (e.g., model exploratory, data poisoning attacks) that aim to sabotage training process. Thus, the required precautions should be taken during the ML process to prevent such attacks. We also believe adversarial machine learning can be adapted to Simian Army \citenum{tseitlin_antifragile_nodate} approach in the context of attack generation.

\subsection{Redundancy and Resilience}

Among the characteristics that determine the overall security of an ICPS, the most overlooked one is a redundancy (see \cref{section:characteristics}) that also directly contributes to the resilience of the system. The German Steel Mill incident \citenum{lee2014german} is such an example of to lack of redundancy where the incident could have been prevented if there was an additional system to shut down the furnace. Regarding resilience, imagine the rotating speed of the fan is altered by an adversary to cause a fire in the manufacturing line. If the fan can go back to a normal state without causing fire we can define that system as resilient. In industrial environments, only the most significant elements (i.e., electricity) are considered from these perspectives where cyber-physical edge security mechanisms subject to replacement when they fail which makes the ICPS vulnerable during the replacement process. Deploying an alternative system for each process is not a feasible option (in most situations, the cost will be too high) due to the heterogeneity of an industrial environment. The same is also applied to industrial networks. As we mentioned in \cref{section:systemdefinitions}, ICPSs may benefit from a highly interconnected network that also increases the attack surface (see \citenum{lindsey_odonnell_post-ransomware_2020} for real-life incident example). This case becomes more significant when there is inadequate boundary protection (see \cref{section:countermeasures}) as we have also seen from the vulnerability assessment reports \citenum{noauthor_ics-cert_2016, noauthor_2020_2020,noauthor_nccicics-cert_2015}. Thus, two main challenges arise: (i) how can we determine the industrial assets that require back-up systems, (ii) how should we implement them. The risk assessment has to be done to analyze available options. Regarding ICPS edge resources, we may question the abandonment of air-gapping policy in ICPSs. Deploying supervision/security mechanisms to the same network allows the adversary to bypass them by manipulating their outputs forwarded to the control center \citenum{chen_stuxnet_2010}. Thus, deploying edge security mechanisms on an air-gapped secondary network seems like a promising further research topic.

\subsection{Testbeds and Synthetic Datasets}

The survey revealed that most of the proposals targeting the top ten identified industrial weaknesses \citenum{noauthor_ics-cert_2016} utilize benchmarking tests, simulations, or proof of concepts to evaluate their proposals. To achieve the most realistic results, real-life datasets and testbeds are required. This is a well-known challenge for industrial studies that exist from the beginning. Experimenting on real and active ICPS is out of the question due to the chance of possible disruption in CIs. Hence, aiming for the utilization of the most realistic testbeds and synthetic datasets \citenum{belenko2018synthetic, vanerio2017ensemble} that benefit from being privacy-free (the reason behind the lack of efficient publicly available datasets) and safe to gather is the most feasible option. 
We believe the development of advanced industrial simulation environments \citenum{airbus} that can generate robust synthetic datasets to be utilized with machine learning models is one of the significant future directions to be considered. To generate such a dataset where the privacy of the data is out of concern that contains both malicious and normal traffic, either we can conduct attacks (a similar approach to the Simian Army \citenum{tseitlin_antifragile_nodate} may be applied) or deploy honeypot to the ICPS testbed network. However, achieving big data is still a challenge with in terms of generating synthetic datasets. On the other hand,  realistic evaluation becomes more challenging if we consider the studies regarding industrial security policies where the efficiency of the policy is mostly determined by the aftermath of the real attacks. Possible directions include simulating real-life incidents based on the choices of developed policies and the development of an evaluation framework that differs according to industrial environments.

\section{Conclusions}
\label{section:sec6}

The industrial cyber-physical systems (ICPS) are adopting new communication, computation based technologies, and becoming more interconnected, heterogeneous, and dynamic. Even though ICPS benefits from this rapid integration, providing cybersecurity is becoming a primary concern due to increased attack surface. In this paper, we reviewed the overall ICPS cybersecurity to understand what the current challenges are and how they are treated by the academia. We analyzed the ICPS architecture by defining its components and emphasizing the unique characteristic of OT systems. We provided an analysis of ICPS communication protocols. We proposed an adaptive attack taxonomy then evaluated real-life ICPS cyber incidents. We analyzed the latest trends on ICPS edge security. Then, we surveyed the growing ICPS cybersecurity literature to determine how academia approaches against ICPS vulnerabilities. We evaluated studies that propose ICPS security mechanisms based on the evaluation metrics that aim for continuity. We argued about the datasets, testbeds, machine learning techniques, security policies that will shape the future of ICPS security. In all papers we surveyed, no paper proposes a framework that explains the relationship with complementary industrial systems. The less utilization of realistic testbeds, lack of up-to-date datasets, and evaluation framework to compare AI/ML techniques are also among major challenges. Besides, the most common vulnerabilities are either due to weak boundary protection or lack of enforcement of well-designed security policy. We hope that our review and suggestions will motivate further research while closing the gap between academia and industry in this field and lead unified studies that focus on adaptive security mechanisms based on strong policies.


\end{document}